\documentclass[twocolumn]{aastex62}
\usepackage{longtable}
\usepackage{threeparttablex}
\usepackage{footmisc}
\usepackage{color}

\submitjournal{AAS Journals}

\shorttitle{SOAR TESS survey}
\shortauthors{Ziegler et al.}

\begin{document}

\title{SOAR TESS Survey. I: Sculpting of TESS planetary systems by stellar companions}

\correspondingauthor{Carl Ziegler}
\email{carl.ziegler@dunlap.utoronto.ca}

\author[0000-0002-0619-7639]{Carl Ziegler}
\affil{Dunlap Institute for Astronomy and Astrophysics, University of Toronto, 50 St. George Street, Toronto, Ontario M5S 3H4, Canada}

\author{Andrei Tokovinin}
\affiliation{Cerro Tololo Inter-American Observatory, Casilla 603, La Serena, Chile} 

\author{C\'{e}sar Brice\~{n}o}
\affiliation{Cerro Tololo Inter-American Observatory, Casilla 603, La Serena, Chile} 

\author{James Mang}
\affiliation{Department of Astronomy, University of California, Berkeley, 510 Campbell Hall, Berkeley, CA 94720, USA}

\author{Nicholas Law}
\affiliation{Department of Physics and Astronomy, The University of North Carolina at Chapel Hill, Chapel Hill, NC 27599-3255, USA}

\author[0000-0003-3654-1602]{Andrew W. Mann}%
\affiliation{Department of Physics and Astronomy, The University of North Carolina at Chapel Hill, Chapel Hill, NC 27599-3255, USA}

\begin{abstract}

TESS is finding transiting planet candidates around bright, nearby stars across the entire sky. The large field-of-view, however, results in low spatial resolution, therefore multiple stars contribute to almost every TESS light curve. High-angular resolution imaging can detect the previously unknown companions to planetary candidate hosts that dilute the transit depths, lead to host star ambiguity, and in some cases are the source of false-positive transit signals. We use speckle imaging on SOAR to search for companions to 542 TESS planet candidate hosts in the Southern sky. We provide correction factors for the 117 systems with resolved companions due to photometric contamination.  The contamination in TESS due to close binaries is similar to that found in surveys of \textit{Kepler} planet candidates. For the solar-type population, we find a deep deficit of close binary systems with projected stellar separations less than 100 AU among planet candidate hosts (44 observed binaries compared to 124 expected based on field binary statistics). The close binary suppression among TESS planet candidate hosts is similar to that seen for the more distant \textit{Kepler} population. We also find a large surplus of the TESS planet candidates in wide binary systems, detected in both SOAR and \textit{Gaia} DR2 (119 observed binaries compared to 77 expected). These wide binaries host almost exclusively giant planets, however, suggesting orbital migration, caused by perturbations from the stellar companion, may lead to planet-planet scattering and suppress the population of small planets in wide binaries.  Both trends are also apparent in the M-dwarf planet candidate hosts.

\end{abstract}


\section{Introduction} \label{sec:intro}
Over a decade-long primary and extended missions, the \textit{Kepler} telescope \citep{borucki10} detected the majority of known exoplanets.  \textit{Kepler} probed the large \citep{howard12} and diverse \citep{lissauer11, welsh12} Galactic population of planetary systems, but only looked at relatively small regions of the sky at a time. Also, the observed stars in the prime mission were generally too distant for precision follow-up observations.

Beginning in 2018 July, the Transiting Exoplanet Survey Satellite (TESS, \citealt{tess}) searched the Southern sky for exoplanets around nearby stars. These typically bright targets are accessible to measurements of planet masses with precision radial velocity and atmospheric characterization of planets with transmission spectroscopy. Several planets around bright stars have been confirmed \citep{huang18, dragomir19, vanderburg19}, with hundreds of additional candidates awaiting follow-up observations.\footnote{\url{https://tess.mit.edu/alerts/}}

The simultaneous field-of-view required for TESS is large, approximately 6\% of the entire sky, which it covers with a relatively coarse pixel scale of 21$\arcsec$ px$^{-1}$. For reference, each TESS pixel subtends a region of the sky approximately 25$\times$ that of each \textit{Kepler} pixel. Each TESS pixel observes the flux made up of the blended contributions of multiple sources. The TESS Input Catalog (TIC, \citealt{ticold, tic}) determined the contamination from known point sources in two catalogs (APASS and 2MASS) likely to be in the TESS aperture for 3.8 million stars. While these catalogs limiting magnitudes are relatively faint compared to the typically bright TESS targets (T$_{mag,limiting}\sim$17-19), they are based on seeing-limited observations and are not sensitive to binaries with separations less than 1-2\arcsec. Likewise, Gaia DR2 generally does not recover binaries with separations of less than $\sim$0\farcs7, in particular nearby bright stars \citep{roboaogaia}. 

High-resolution imaging has proven to be critical to confirm and characterize transiting planet candidates. Half of solar-type stars \citep{raghavan10} and a quarter of M-dwarfs \citep{winters19} are found in multiple systems. The maximum of the distribution of orbital separations ($\sim$50 AU for a solar-type binary and 20 AU for M-dwarfs), at a typical distance to a TESS host star of approximately 100 pc, peaks at angular separations of 0\farcs2 to 0\farcs5, accessible only by high angular resolution imaging. Therefore, most contamination from binary systems is not accounted for in the TIC. 

For \textit{Kepler}, the planet radius estimates for stars with detected companions increased by a factor of 1.6 on average \citep{ciardi15, ziegler18a}. In many cases, the identity of the planetary host star may be ambiguous, leading to two different possible radius estimates based on either scenario. In addition, the absence of companions can be used to rule out many common false positive scenarios, allowing planets to be statistically validated \citep{torres11, morton11}.

The number of planet candidates detected by \textit{Kepler} and currently by TESS outstrips the resources available on conventional high-resolution instruments, such as laser-guide star adaptive optics.\footnote{For example, the acquisition time of Keck-AO averages 9 minutes \citep{wizinowich06}, so over a hundred nights would be needed to observe the \textit{Kepler} planet candidates.} A useful strategy with \textit{Kepler} was to perform a broad survey with a visible-light high-resolution instrument on a moderately sized telescope. These observations are able to find the majority of both physically bound companions and low-contrast asterisms that significantly alter the radius estimate of the planet candidate. The Robo-AO \textit{Kepler} survey \citep{law14, baranec16, ziegler17} observed 3857 planet candidate host stars with a laser-assisted adaptive optics (AO) robotic instrument on a 2-m class telescope \citep{baranec14}. The discoveries of companions corrected the radius estimates for 814 planet candidates \citep{ziegler18a}, contributed to the validation of over a thousand \textit{Kepler} planets \citep{morton16}, and informed future observations with large-aperture telescopes. Six years after the \textit{Kepler} prime mission ended, over half of the \textit{Kepler} planet candidates have only been imaged at high-resolution by Robo-AO.

The TESS sample can largely be covered by speckle interferometry, due to host stars that are on average several magnitudes brighter than \textit{Kepler}. Speckle interferometry on SOAR has been developed and optimized over the past decade, and at present can image up to 300 targets a night with diffraction-limited resolution \citep{tokovinin18}. \citet{mann19} found the astrometric precision of SOAR speckle imaging to be among the best compared to similar non-fixed high-resolution instruments. Imaging in Cousins-I band, at the center of the TESS bandpass, the dilution in the TESS light curves due to detected companions can be accurately determined to correct radius estimates.

\begin{figure*}
    \centering
    \includegraphics[scale=0.65]{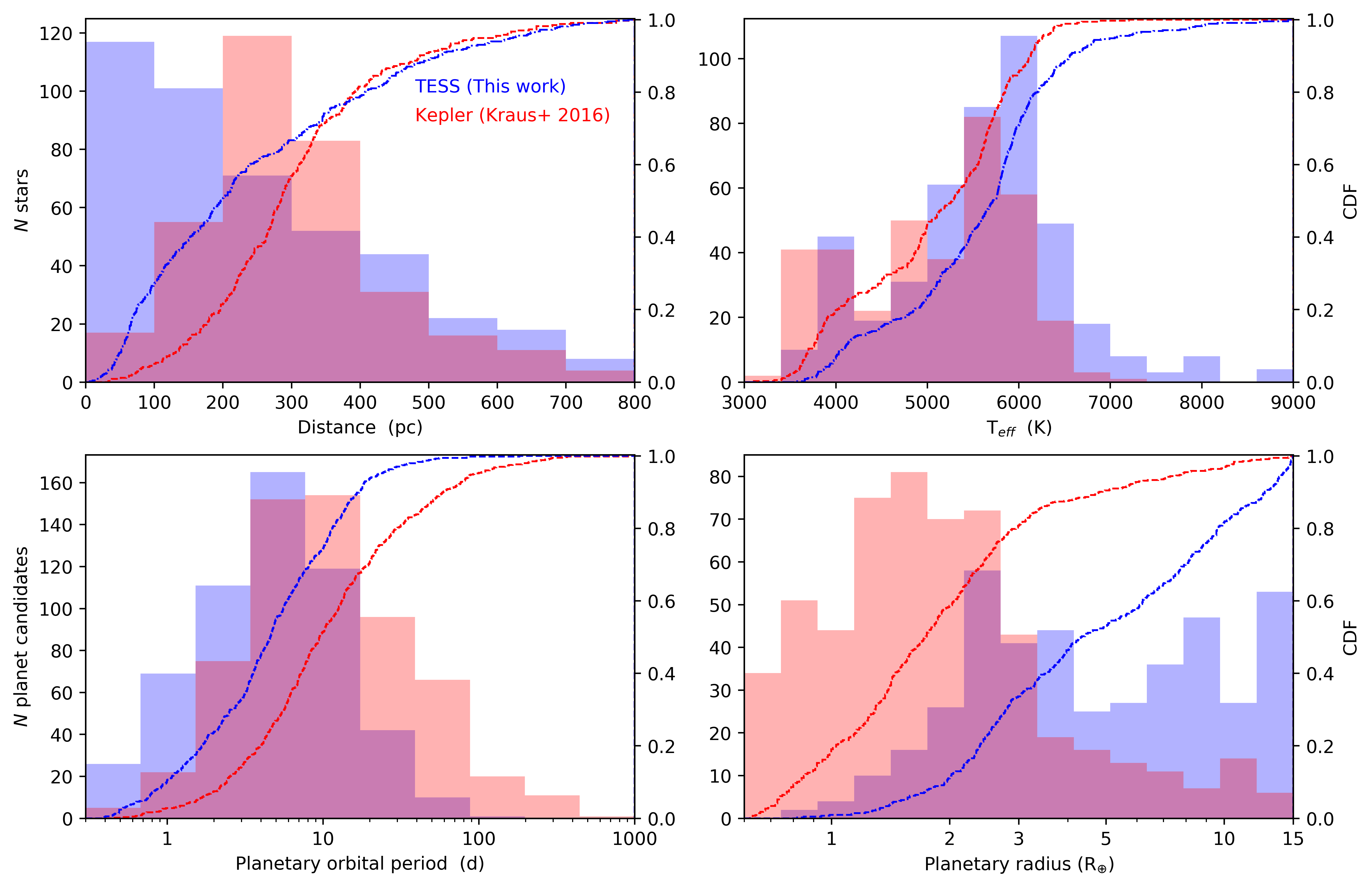}
    \caption{The properties of the 542 TESS planet candidate hosts observed by SOAR in this survey, presented as a binned histogram with an overplotted cumulative density function. For comparison, the properties of the 382 \textit{Kepler} planet candidate hosts observed in \citet{kraus16} are also plotted. In general, the TESS planets are closer to the solar system, have slightly hotter hosts, orbit with shorter periods, and are larger than the \textit{Kepler} planets targeted by the \citet{kraus16} survey.}
    \label{fig:histogram}
\end{figure*}

Theory suggests the presence of a nearby stellar companion can have a dire impact on the formation of circumstellar planetary systems: stirring planetesimals \citep{quintana07}, perturbing orbits resulting in high eccentricity tidal migration \citep{naoz12}, truncating protoplanetary discs and shortening their lifetime \citep{jangcondell08, kraus12}, and increasing photoevaporation \citep{alexander12}. \citet{ngo16} found that hot Jupiters have significantly fewer companions at close separations compared to field stars, but many more at wider separations. This suggests these hot Jupiters may have been driven inwards to their present few-day orbits by perturbations from the companion stars. Bolstering this interpretation, the Robo-AO \textit{Kepler} survey found that hot Jupiters were significantly more likely to be found in binary systems than other types of planets \citep{ziegler18b}. \citet{deacon16}, however, found no difference in the wide ($\rho>3000 AU)$ binary rate between transiting planet hosts and field stars in the \textit{Kepler} field.

\citet{kraus16}, observing 382 \textit{Kepler} planet candidates with adaptive optics on Keck, found a dearth of \textit{Kepler} planets in close binary systems. This deficit was modeled as a binary suppression factor of 0.34  at separations below approximately 47 AU. Extrapolating this out suggests that one-fifth of the solar-type stars in the galaxy can not host planets due to the influence of a stellar companion. It is unclear, however, if the survival of planets in close binary systems is random, or a result of other factors, such as the binary eccentricity or the mutual inclination to the planetary system of the binary system. The detection image provides only an instantaneous projected separation, $s$. Further monitoring is needed to determine the true orbital parameters that could provide insight into how some planetary systems form and survive in this harsh environment.

The TESS planet candidate hosts are relatively nearby; on average, less than half the distance as the \textit{Kepler} hosts, based on the TIC distance estimates \citep{tic}. The 4.1m SOAR telescope can, therefore, detect companions at solar-system separations ($s$=10-50 AU) to the vast majority of TESS targets. Evidence of suppression in the binary rate for TESS planet candidates in this regime would serve both as an independent validation of the ruinous effect binaries have on planetary systems, and, since the TESS planets are spread over the entire sky, confirmation of the effect in a more representative sample of the Galactic planetary population. Indeed, \citet{matson18} did not detect binary suppression in a sample of K2 stars, which are spread in fields across the ecliptic plane. The authors note the non-detection is tenuous, however, and more high-resolution observations of exoplanet hosts are needed.

We begin in Section \ref{sec:observations} by detailing our observations and data analysis. We present the results of the survey in Section \ref{sec:results}, and explore the impact binaries have on the TESS planets in Section \ref{sec:analysis}. We discuss the results further in Section \ref{sec:discussion}. Finally, we conclude in Section \ref{sec:conclusions}.

\section{Observations and Analysis}\label{sec:observations}

\subsection{Target selection}

The hosts of TESS planet candidates (TESS objects of interest, or TOIs) were selected from the data releases available online at the TESS data release portal.\footnote{\url{https://tev.mit.edu/toi/}, account required for access.} Faint stars (typically, T$_{mag}>$13 mag) that are not well suited for speckle observations were not targeted; this limit reduces the number of late-type stars that are observed in this survey. Previously confirmed planet hosts, primarily from the WASP \citep{wasp} and HATS \citep{hats} surveys, were excluded from the target selection as these systems have been heavily studied in the past \citep[e.g.,][]{ngo15, evans16, evans18}. Seventeen community detected TESS planet candidates\footnote{From CTOI list, available at \url{https://exofop.ipac.caltech.edu/tess/view_ctoi.php}} were also observed but were not used in the subsequent analysis in this work. To increase observing efficiency, target acquisition was improved using precise target coordinates, determined for each night with proper motions from Gaia DR2 \citep{gaia}, when available, and from the TIC \citep{tic} otherwise. As previously noted in \citet{arenou18}, we find that many targets with only two-parameter astrometric solutions in Gaia DR2 are actually close binaries.

The properties of the host stars and planet candidates observed are plotted in Figure \ref{fig:histogram}.

\subsection{SOAR observations and data reduction}

We observed 542 TESS planet candidate hosts with the high-resolution camera (HRCam) imager on the 4.1-m Southern Astrophysical Research (SOAR) telescope over seven nights in 2018-2019. The observation procedure and data reduction are described in \citet{tokovinin18}. Briefly, each observation consists of 400 frames split into two data cubtes, typically consisting of a 200$\times$200 binned pixels region of interest centered on the target star (6\farcs3 on a side at the pixel scale of 0\farcs01575 and 2$\times$2 binning), taken in approximately 11 s with an Andor iXon-888 camera. The resulting data cube is processed by a custom IDL script, which computes the power spectrum, in which a resolved multiple stellar system will appear as characteristic fringes. Binary parameters (separation, position angle, and magnitude difference) are determined from modeling the power spectrum. Secondary stars will appear as mirrored peaks in the speckle auto-correlation function (ACF), the Fourier transform of the power spectrum, at the separation and position angle of the companion. To remove the 180-degree ambiguity inherent to the classical speckle interferometry, our pipeline also computes the shift-and-add (SAA) images, centered on the brightest pixel in each frame (this is sometimes called ``lucky imaging"). Relatively bright binaries with $\Delta m > 0.5$ mag often have their companions visible in the SAA images, allowing us to select the correct quadrant (these measurements are marked by the flag `q' in Table \ref{tab:whitelist} in the Appendix). In all other cases, the position angles are determined with a 180\degr ambiguity. Figure 4 in \citet{tokovinin18} gives an example of typical speckle data, including the SAA image.  Observations were taken in the $I$-band ($\lambda_{cen}$=824 nm, $\Delta\lambda$=170 nm), which is approximately centered on the TESS bandpass. Four resolved systems (TOI-123, 131, 138, and 146) were also imaged in $V$-band in preparation for a future association analysis. 

The detection limits are estimated from the RMS fluctuations $\sigma$ of the ACF computed in annular zones of increasing radii; peaks exceeding the 5$\sigma$ level are assumed to be detectable. This has been verified by injecting simulated binary companions in the real ACFs (see e.g. Fig. 9 in \citet{tokovinin10}). Moreover, for each target we require detection of the companion in {\em both} data cubes. Our procedure practically excludes false detections at separations larger than $\sim$0\farcs1. At closer separations, however, the faint and persistent ACF details of instrumental origin, such as vibration or the optical ghosts described in \citet{tokovinin18}, can mimic real close binaries with a large $\Delta m$. We inter-compare the ACFs and power spectra of sequentially observed targets to identify such artifacts. Finally, the estimated detection limits are verified by comparing with $(\rho, \Delta m)$ of actually measured companions, and a good agreement is always found.

The pixel scale and orientation are calibrated by observing several wide binaries with a well-modeled motion, as explained in Section 4.2 of \citet{tokovinin18}. The RMS agreement between the measured calibrator positions and their models is typically from 1 to 3\,mas in separation and better than 0\fdg2 in angle. The pixel scale is therefore established with an accuracy of better than 0.5\%. The calibration of HRCam was checked by comparing the positions of wider calibrators predicted by their models for 2015.5 with their relative positions in Gaia DR2.

\begin{figure}
    \centering
    \includegraphics[scale=0.57]{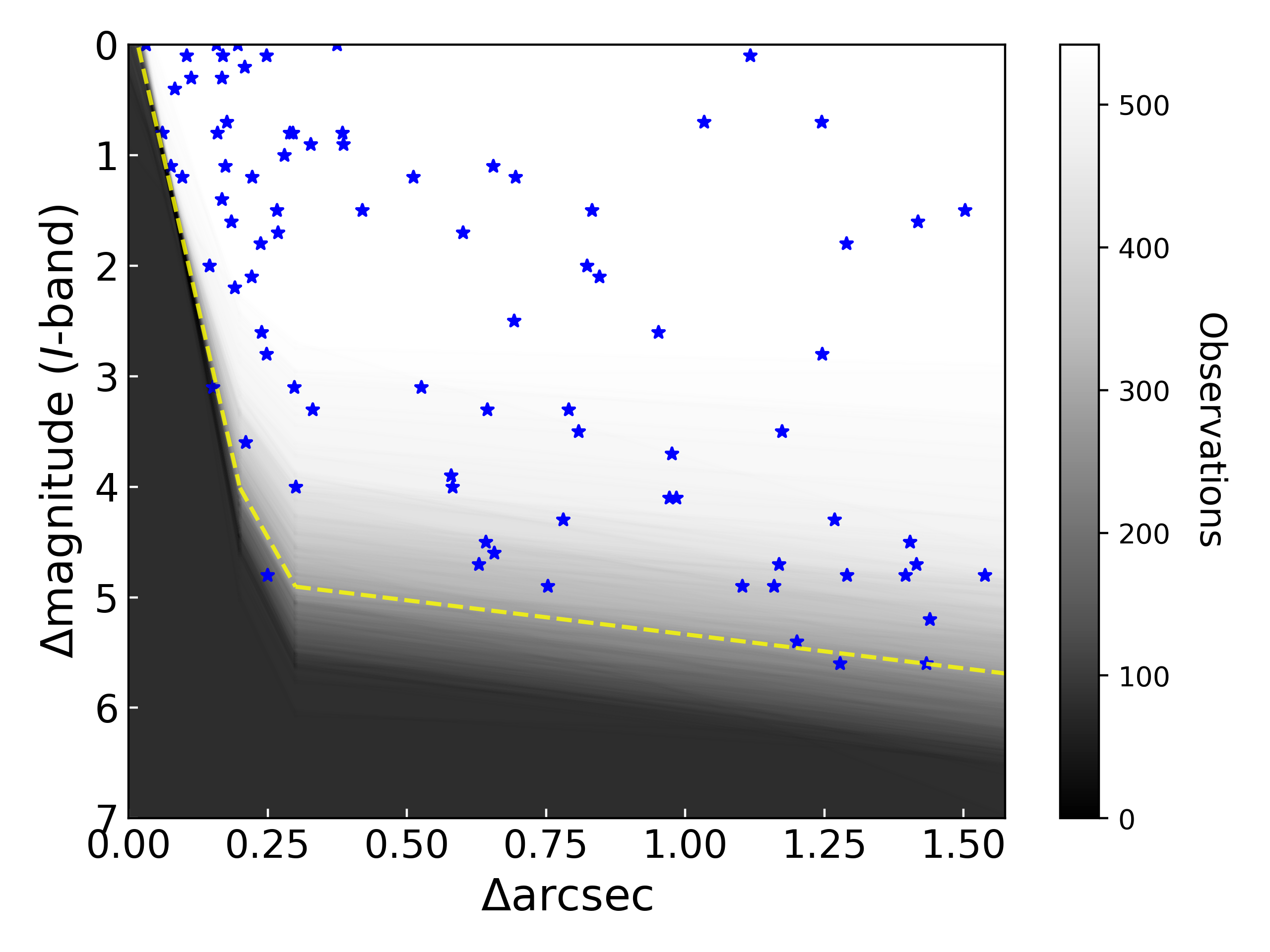}
    \caption{Close companions ($\rho <$1\farcs55) to TESS planet candidate hosts detected by SOAR speckle imaging, in terms of their $I$-band magnitude difference and separation from the primary star. The average 5-$\sigma$ detection limits of the observations are plotted, trending from black (no observations are sensitive to binaries with that combination of separation and contrast) to white (all 542 observations are sensitive to that combination). The yellow dashed line shows the median sensitivity for the survey.}
    \label{fig:binariesplot}
\end{figure}

We detail the observations in Table \ref{tab:whitelist} in the Appendix. The cumulative 5-$\sigma$ detection sensitivities are plotted in Figure \ref{fig:binariesplot}. 

\subsection{Planet radius corrections}\label{sec:radiuscorrections}

The additional flux from a nearby star will dilute the transit depth in the TESS light curves, resulting in an underestimated radius for the planet candidate. We compute correction factors to the radius estimates derived from the TESS light curves for two scenarios: 1) the planet orbits the target star; and 2) the planet orbits the secondary star which is bound to the primary star.\footnote{A third scenario, in which the secondary star is unbound to the primary star, is unconstrained without color information. In future papers, we will use multi-band speckle imaging with SOAR to extend our analysis to the unbound scenario.}

For the first scenario, we use the relation from \citet{law14} to derive a radius correction factor,
\begin{equation}\label{eq:1}
X_{prim}=\frac{R_{p,A}}{R_{p,0}}=\sqrt{\frac{1}{F_{A}}}
\end{equation}
where R$_{p,A}$ is the corrected radius of the planet orbiting the primary star, R$_{p,0}$ is the original planetary radius estimate based on the diluted transit signal, and F$_{A}$ is the fraction of flux within the aperture from the primary star.

For the case where the planet candidate is bound to the secondary star, we use the relation for the radius correction factor,
\begin{equation}\label{eq:2}
X_{sec}=\frac{R_{p,B}}{R_{p,0}}=\frac{R_{B}}{R_{A}}\sqrt{\frac{1}{F_{B}}}
\end{equation}
where R$_{p,B}$ is the corrected radius of the planet orbiting the secondary star bound to the primary star, R$_{B}$ and R$_{A}$ are the stellar radii of the secondary and primary star, respectively, and F$_{B}$ is the fraction of flux within the aperture from the secondary star.

We use the stellar radius estimates from the TIC \citep{tic} when available for the primary stars. The radii of secondary companions in the scenario where they are bound to the target star were estimated using the observed contrast ratio in the \textit{TESS} band (approximated using the $I$ bandpass of SOAR) and finding the radius of an appropriately fainter star within the Dartmouth stellar models \citep{dotter08}.

The TIC includes a contamination ratio that takes into account stars within 10 TESS pixels of the target. This includes stars typically down to the limiting magnitude of the 2MASS \citep{skrutskie06} and APASS \citep{Henden09} catalogs ($T\sim17-19$). Using the list of detected close binaries to \textit{TESS} planet candidate hosts and their binary parameters, a custom Python script crossmatched each of their coordinates to stars in the TIC catalog. We find 31 stars in the TIC had similar positions relative to the primary as was found in SOAR imaging ($\Delta\rho<0\farcs5$ and $\Delta \theta <$20\degr, or $[\Delta \theta\pm180^{o}]<$20\degr). The properties of these systems are available in Table \ref{tab:ticbinaries} in the Appendix. One notable resolved binary in our survey is the pair TOI-658 and TOI-659. The magnitude differences in the TESS bandpass for wide binaries in the TIC are generally similar to that measured from the SOAR observations, supporting our use of $I$-band observations as a proxy for the TESS observations. A similar crossmatch was performed with Gaia DR2 \citep{gaia}, yielding 38 matches to SOAR detected companions. These companions were all widely separated ($\rho>1\arcsec$). The separations measured by SOAR have a mean and median difference of 15 and 6 mas, respectively, compared to those reported in Gaia DR2. The average difference in magnitude differences between SOAR and Gaia DR2 is 0.16 mag, and is likely due to the different passbands. The properties of these systems are available in Table \ref{tab:gaiamatches} in the Appendix.

We provide a correction factor for hosts, as in some cases the crossmatch between the TIC and the SOAR binary is ambiguous, however, we caution that the correction should be used judiciously. For all other systems, the contamination ratios reported should be used in addition to the TIC contamination ratio. In practice, the reported radius estimates of TESS planet candidates on the TESS data release portal and ExoFOP typically take into account flux contamination. The additional correction due to binaries detected by SOAR is the product of the original radius estimate and the radius correction factor reported in this work.

\subsection{Physical association of companions} \label{sec:unboundedstars}

\begin{figure}
    \centering
    \includegraphics[scale=0.55]{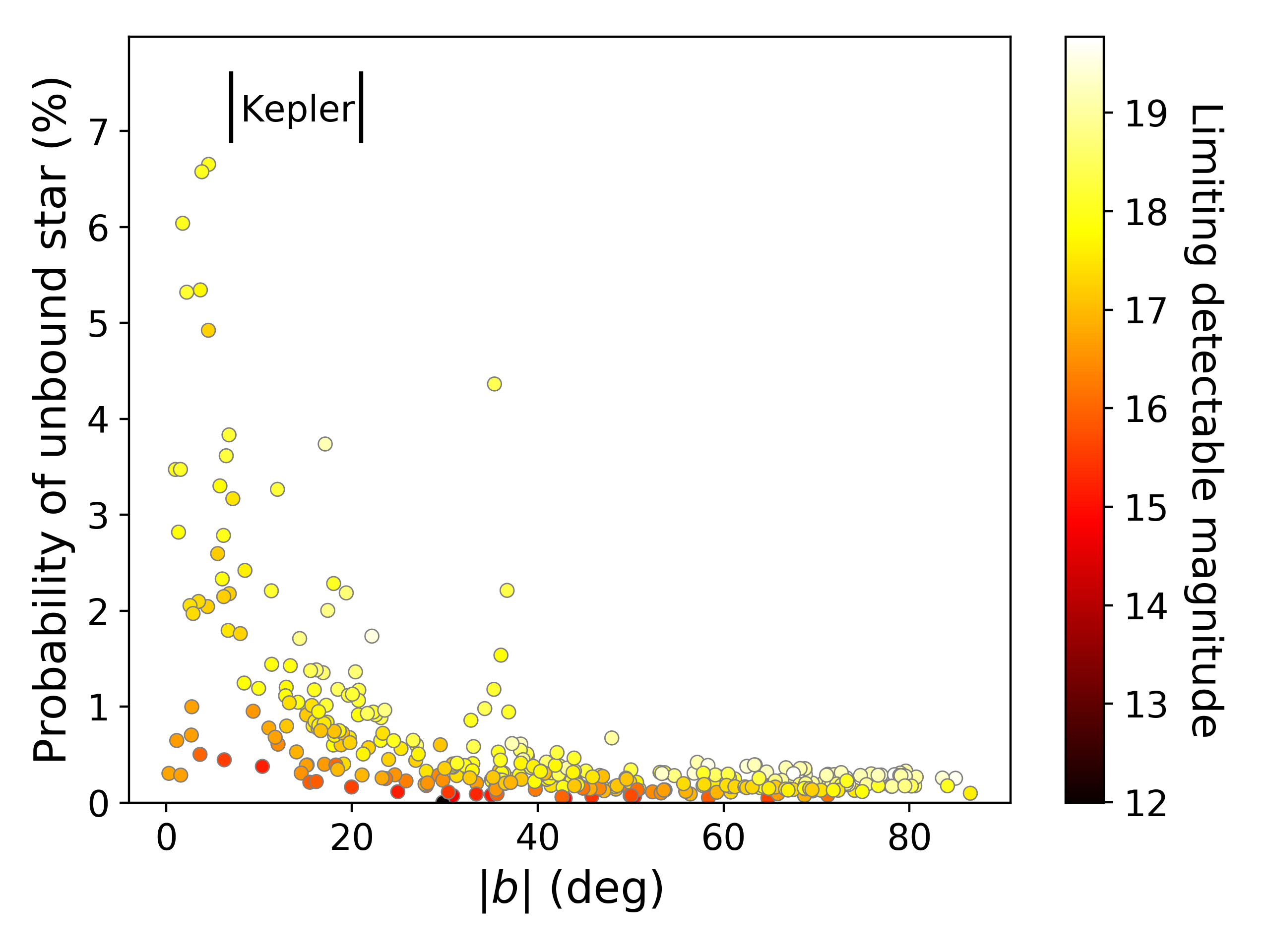}
    \caption{The probability of an unbound star being detected within the search radius of 3\farcs15 for each observed TESS planet candidate as a function of Galactic latitude, based on $Gaia$ DR2 stellar counts. Each target is colored by the limiting magnitude of detectable nearby stars in the SOAR speckle observations. The latitude range of the primary \textit{Kepler} mission is indicated for reference. We expect approximately three unbound stars in total to be detected near the TESS planet candidates.}
    \label{fig:unboundstars}
\end{figure}

A relatively large number of companions detected near \textit{Kepler} planet candidates were unassociated, especially at separations greater than 1$\arcsec$ \citep{horch14, ziegler18b}. The TESS targets are spread across the sky, in regions of low and high stellar density, but generally at higher Galactic latitudes than the \textit{Kepler} field. In addition, the targets are typically bright (T$_{mag}<$12 mag), and subsequently the detectable stars near them are several magnitude brighter than the \textit{Kepler} stars, given approximately equal contrast sensitivity. It is likely then that the number of detected field stars will be reduced in the TESS sample.

We use the stellar densities in the region of sky around each target in Gaia DR2 \citet{gaia} to estimate the likelihood of a field star being detected near a TESS star. Gaia DR2 is essentially complete for sources down to G=17 mag, and down to G=19 mag in non-crowded fields \citep{gaia}. For our typical target with $I=12$ mag and a contrast sensitivity of 5 mag, the faintest detectable companions will have $I \approx 17$ mag, comparable to the \textit{Gaia} completeness limit.

For each target, we begin by performing a cone search in DR2 to search for all stars within 0.5$^{\circ}$. The cone search was done using the Astropy affiliated package astroquery \citep{astropy:2018} to search the Gaia DR2 catalog hosted by the Barbara A. Mikulski Archive for Space Telescopes (MAST).  We use the number of sources in DR2 within the 0.5$^{\circ}$ circular field to determine the source density (i.e., number of sources per arcsec$^{2}$) as a function of $G$ magnitude, $\rho_{source}[G]$. For each target, we determine the limiting magnitude of detectable secondary stars ($G_{limit}$) within separation increments of 0\farcs5. We then use these limiting magnitudes and the DR2 source density ($\rho_{source}[G<G_{limit}]$) to determine the probability of detecting an unassociated field star within that separation range (e.g., 1\arcsec to 1.5\arcsec). The cumulative probability for all separations, out to 3\farcs15, provides the number of field stars we can expect to find near that target (see Figure \ref{fig:unboundstars}). Typically, targets near the galactic plane and near the Small and Large Magellanic clouds are far more likely to have a field companion due to the high local source density.

We perform a Monte Carlo analysis using these probabilities and the distribution of contrast sensitivity to simulate 10$^{4}$ surveys and find that on average, we should detect 3.2$\pm0.5$ field stars within 3\farcs15 of the observed TESS targets. The field companions are nearly always high-contrast (with a large $\Delta I$). We, therefore, expect the impact on the subsequent analysis due to unassociated asterisms of field stars to be negligible.

\begin{figure}
    \centering
    \includegraphics[scale=0.52]{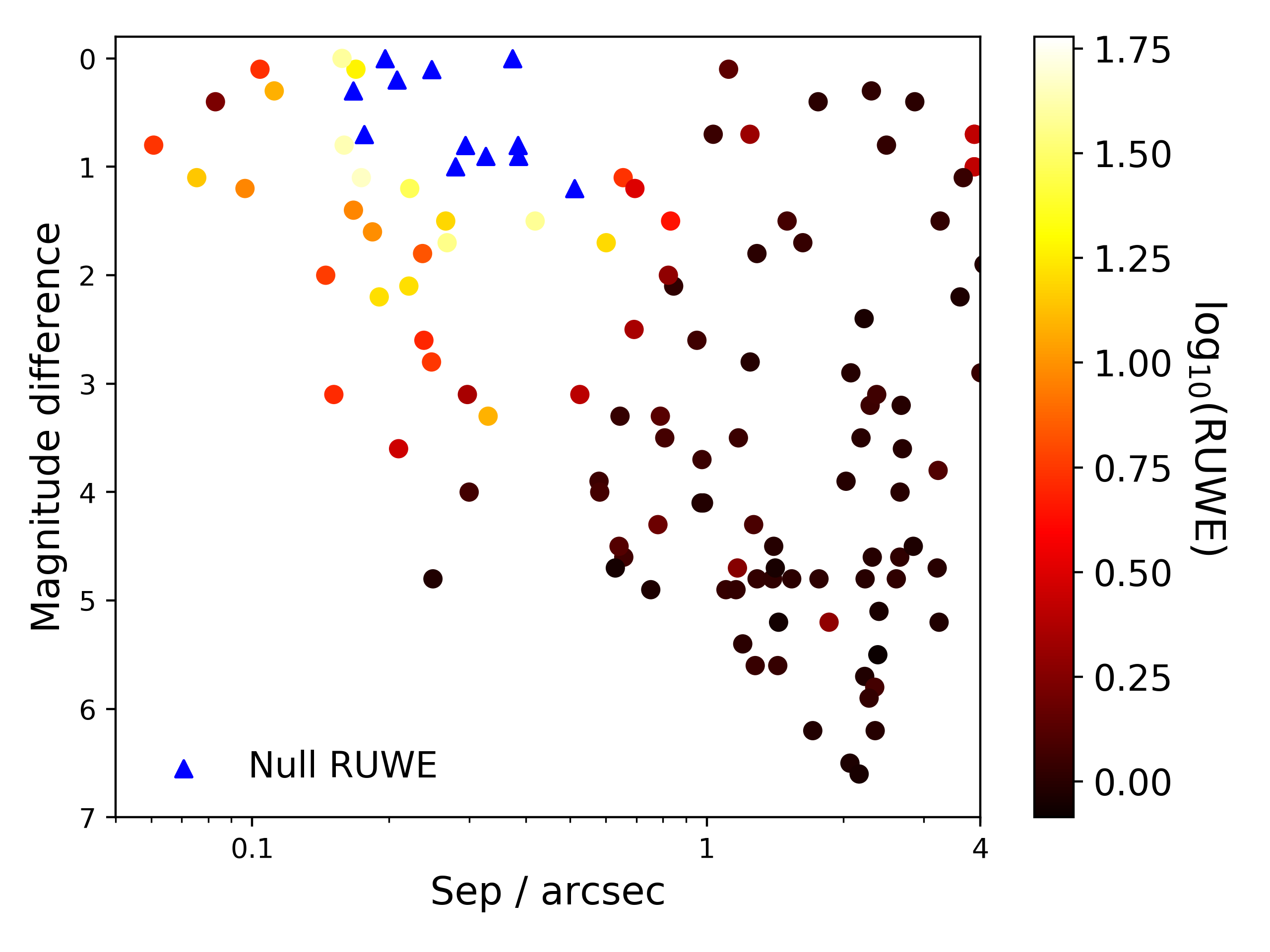}
    \caption{The properties of companions to TESS planet candidate hosts resolved by SOAR, colored by the logarithm of each system's Gaia DR2 RUWE value, a metric quantifying the quality of the Gaia astrometric solution. Companions in systems with null RUWE values are indicated by a blue triangle. While Gaia does not resolve close binaries ($\rho<0\farcs7$), the high or null RUWE value may be used to infer the existence of a companion star. This method is successful with an accuracy of 86$\%$ in our survey.}
    \label{fig:gaiaruwe}
\end{figure}

\section{Results}\label{sec:results}

We detected 88 and 126 companions within 1\farcs5 and 3$\arcsec$ of 84 and 117 TESS planet candidate hosts, respectively, out of 542 observed with speckle imaging on SOAR. This implies a companion rate within 1\farcs5 and 3$\arcsec$ for TESS planet candidates of 16.2$\pm$1.7$\%$ and 23.2$\pm$2.0$\%$, respectively. The properties of the detected companions are plotted in Figure \ref{fig:binariesplot}, along with the average detection sensitivities from all observations which are detailed in Table \ref{tab:whitelist} in the Appendix. We include the radius correction factors for planets in these systems, whether they orbit the primary or secondary star. The auto-correlation functions of resolved systems showing the position of the companions are shown in Figures \ref{fig:grid1}-\ref{fig:grid4} in the Appendix.

The results of each night's observations were processed within a week and posted on the TESS Exoplanet Follow-up Observing Program (ExoFOP) webpage\footnote{\url{https://exofop.ipac.caltech.edu/tess/}} to aid in confirmation of the planet candidates. Several studies \citep{vanderburg19, rodriguez19, quinn19, espinoza19, jones19} have used the SOAR speckle results to confirm early TESS planets.

\subsection{High-order multiples}

Nine targets are resolved triples with two companions in the SOAR images. For TOIs 378, 183, 487, 952, and 909, both companions are paired to the main star. In contrast, close pairs in TOI 455, 697, and 612 belong to the secondary components. Finally, in TOI 141 both separations are comparable, 1\farcs2 and 0.44\arcsec. For the resolved triples, the positions and magnitude differences refer to pairs of stars (e.g. Aa,B and Aa,Ab) as indicated in Column 3 of Table \ref{tab:whitelist} in the Appendix, not the merged inner pairs like A,B.

\tabcolsep=0.12cm
\begin{tiny}

\begin{ThreePartTable}
\begin{TableNotes}
\footnotesize

\end{TableNotes}

\begin{longtable*}{ccccccccccc}
\caption{Nearby stars detected by SOAR to TESS planet candidate hosts \label{tab:binaries}}\\
\hline
\hline
\noalign{\vskip 3pt}  
\text{TOI} & \text{Separation} & \text{P.A.} & \text{Contrast} & \text{T$_{eff}$} & \text{Distance} & \text{Proj. Sep.}\tnote{a} & \multicolumn{2}{c}{Radius correction factor} & Prev. & WDS DD\\ 
 & \text{(\arcsec)} & (deg) & \text{($I$-band)} & \text{(K)} &  \text{(pc)} & \text{(AU)} & \text{(primary host)} & \text{(secondary host)} & det.? &  \\ [0.1ex]
\hline
\noalign{\vskip 3pt} 
\endfirsthead

\multicolumn{11}{c}
{\tablename\ \thetable\ -- \textit{Continued}} \\
\hline \hline
\noalign{\vskip 3pt} 
\text{TOI} & \text{Separation} & \text{P.A.} & \text{Contrast} & \text{T$_{eff}$} & \text{Distance} & \text{Proj. Sep.}\tnote{a} & \multicolumn{2}{c}{Radius correction factor} & Prev. & WDS DD\\ 
 & \text{(\arcsec)} & (deg) & \text{($I$-band)} & \text{(K)} &  \text{(pc)} & \text{(AU)} & \text{(primary host)} & \text{(secondary host)} & det.? &  \\ [0.1ex]
\hline
\noalign{\vskip 3pt}  
\endhead
\endfoot
\hline
\endlastfoot
123 & 1.2894 & 294.6 & 1.8 & 6188 & 72 & 92 & 1.091 & 2.404 & 3 & SEE443 \\ 
128 & 2.2195 & 153.8 & 2.4 & 6086 & 67 & 148 & 1.053 & 3.13 & 2 & FIN92 \\ 
131 & 0.0755 & 207.1 & 1.1 & 4174 & 55 & 4 & 1.168 & 1.932 & 1 & \\ 
138 & 0.0964 & 257.0 & 1.2 & 5722 & 128 & 12 & 1.154 & 2.005 & 1 & \\ 
141 & 1.1999 & 305.2 & 5.4 & 5795 & 48 & 57 & 1.003 & 10.881 & 1 & \\ 
141 & 0.4429 & 239.5 & 4.9 & 5795 & 48 & 21 & 1.005 & 8.66 & 1 & \\ 
146 & 0.2896 & 76.1 & 0.8 & 6905 &  &  & 1.216 &  & 1 & \\ 
147 & 2.6583 & 217.6 & 4.6 & 5620 & 820 & 2179 & 1.007 &  & 3 & \\ 
149 & 1.117 & 132.6 & 0.1 & 5914 & 376 & 419 & 1.383 &  & 3 & TDS161\\ 
153 & 0.1691 & 173.6 & 0.1 & 6142 & 436 & 73 & 1.383 & 1.425 & 1 & \\ 
159 & 0.6444 & 17.5 & 3.3 & 6979 & 333 & 214 & 1.024 & 4.594 & 1 & \\ 
161 & 0.1961 & 101.2 & 0.0 & 5388 & 333 & 65 & 1.414 & 1.414 & 1 & \\ 
165 & 2.4855 & 254.5 & 0.8 & 6038 & 193 & 479 & 1.216 & 1.686 & 2 & RST164\\ 
167 & 0.1593 & 306.6 & 0.8 & 5551 & 293 & 46 & 1.216 & 1.776 & 3 & \\ 
172 & 1.1021 & 319.1 & 4.9 & 5911 & 225 & 247 & 1.005 & 9.603 & 1 & \\ 
180 & 1.2682 & 63.7 & 4.3 & 6082 & 270 & 342 & 1.009 & 7.197 & 3 & \\ 
183 & 0.083 & 98.5 & 0.4 & 6270 & 200 & 16 & 1.301 & 1.495 & 1 & \\ 
183 & 3.1806 & 302.0 & 5.1 & 6270 & 200 & 636 & 1.005 & 10.056 & 1 & \\ 
187 & 0.8456 & 104.5 & 2.1 & 6725 & 177 & 149 & 1.07 & 2.797 & 1 & \\ 
188 & 1.2904 & 5.8 & 4.8 & 6340 & 203 & 261 & 1.006 & 8.547 & 1 & \\ 
189 & 0.9716 & 341.7 & 4.1 & 6054 & 177 & 171 & 1.011 & 6.411 & 1 & \\ 
211 & 0.2083 & 258.3 & 0.2 & 5873 & 121 & 25 & 1.353 & 1.338 & 1 & \\ 
224 & 0.0607 & 31.1 & 0.8 & 3689 & 64 & 3 & 1.216 & 1.63 & 1 & \\ 
235 & 0.8326 & 291.6 & 1.5 & 5454 & 104 & 86 & 1.119 & 2.151 & 2 & B2080 \\ 
240 & 2.3656 & 197.6 & 3.1 & 4333 & 74 & 175 & 1.028 & 4.128 & 3 & \\ 
242 & 0.8233 & 164.3 & 2.0 & 6083 & 589 & 484 & 1.076 & 2.66 & 1 & \\ 
245 & 1.6268 & 258.2 & 1.7 & 6208 & 126 & 204 & 1.1 & 2.313 & 2 & B584\\ 
252 & 0.4192 & 311.5 & 1.5 & 5071 & 190 & 79 & 1.119 & 2.105 & 1 & \\ 
253 & 2.6623 & 351.2 & 4.0 & 4020 & 30 & 79 & 1.012 & 5.734 & 3 & \\ 
258 & 2.0749 & 343.2 & 2.9 & 6474 & 561 & 1164 & 1.034 & 3.662 & 3 & \\ 
264 & 0.6413 & 331.5 & 4.5 & 5773 & 422 & 270 & 1.008 & 7.22 & 1 & \\ 
268 & 2.3025 & 51.2 & 0.3 & 5868 & 305 & 702 & 1.326 & 1.373 & 2 & HU1368\\ 
293 & 0.7905 & 0.3 & 3.3 & 5817 & 313 & 247 & 1.024 & 4.22 & 1 & \\ 
295 & 0.9838 & 151.8 & 4.1 & 5663 & 389 & 382 & 1.011 & 6.656 & 1 & \\ 
308 & 0.1576 & 45.2 & 0.0 & 4416 & 201 & 31 & 1.414 &  & 1 & \\ 
309 & 0.3266 & 77.9 & 0.9 & 5312 & 345 & 112 & 1.199 & 1.692 & 1 & \\ 
322 & 0.1761 & 126.5 & 0.7 & 5868 & 277 & 48 & 1.235 & 1.537 & 1 & \\ 
325 & 0.5815 & 222.2 & 4.0 & 4275 & 180 & 104 & 1.012 &  & 1 & \\ 
337 & 0.1451 & 60.8 & 2.0 & 5369 & 292 & 42 & 1.076 & 2.606 & 1 & \\ 
340 & 0.237 & 338.0 & 1.8 & 5655 & 497 & 117 & 1.091 & 2.39 & 1 & \\ 
343 & 2.8471 & 321.0 & 4.5 & 5695 & 438 & 1247 & 1.008 & 7.817 & 3 & \\ 
346 & 1.4329 & 285.3 & 5.6 & 5835 & 745 & 1067 & 1.003 &  & 1 & \\ 
348 & 1.3959 & 40.8 & 4.8 & 5714 & 420 & 586 & 1.006 & 8.958 & 1 & \\ 
364 & 0.3739 & 96.1 & 0.0 & 6219 & 265 & 99 & 1.414 & 1.414 & 1 & \\ 
369 & 0.6949 & 139.4 & 1.2 & 6228 & 126 & 87 & 1.154 &  & 2 & B600\\ 
372 & 0.2386 & 260.6 & 2.6 & 5400 & 340 & 81 & 1.045 & 3.334 & 1 & \\ 
378 & 0.1839 & 330.0 & 1.6 & 5894 & 629 & 115 & 1.109 & 2.316 & 1 & \\ 
378 & 3.3108 & 312.3 & 1.5 & 5894 & 629 & 2082 & 1.119 & 2.232 & 1 & \\ 
379 & 0.031 & 28.8 & 0.0 & 5895 & 227 & 7 & 1.414 & 1.414 & 1 & \\ 
383 & 0.2665 & 333.6 & 1.5 & 5951 & 204 & 54 & 1.119 & 2.118 & 1 & \\ 
386 & 1.1739 & 274.4 & 3.5 & 8100 & 452 & 530 & 1.02 & 5.111 & 3 & \\ 
387 & 2.2878 & 342.9 & 3.2 & 6191 & 193 & 441 & 1.026 & 4.306 & 3 & \\ 
394 & 3.2284 & 220.2 & 3.8 & 6329 & 149 & 481 & 1.015 & 5.583 & 2 & BU529 \\ 
402 & 1.4388 & 233.4 & 5.2 & 5175 & 44 & 63 & 1.004 & 10.825 & 1 & \\ 
405 & 0.3856 & 321.5 & 0.9 & 6138 & 266 & 102 & 1.199 & 1.785 & 1 & \\ 
409 & 3.2443 & 20.4 & 5.2 & 4986 & 53 & 171 & 1.004 & 10.819 & 1 & \\ 
415 & 2.2303 & 89.6 & 4.8 & 6471 & 381 & 849 & 1.006 & 8.547 & 1 & \\ 
422 & 1.4038 & 118.8 & 4.5 & 5823 & 98 & 137 & 1.008 & 7.22 & 1 & \\ 
427 & 2.6122 & 152.9 & 4.8 & 5409 & 145 & 378 & 1.006 & 8.843 & 1 & \\ 
433 & 4.0098 & 324.5 & 2.9 & 8543 & 498 & 1996 & 1.034 & 3.795 & 3 & A3013 \\ 
454 & 3.8682 & 256.7 & 2.2 & 6849 & 78 & 301 & 1.064 & 2.877 & 2 & I56\\ 
455 & 1.0333 & 317.5 & 0.7 & 3562 & 20 & 20 & 1.235 & 1.608 & 2 & RST2292BC\\ 
462 & 0.167 & 196.1 & 0.3 & 5205 & 205 & 34 & 1.326 & 1.497 & 1 & \\ 
463 & 0.8084 & 215.0 & 3.5 & 6497 & 177 & 143 & 1.02 & 4.761 & 1 & \\ 
476 & 0.6568 & 145.4 & 4.6 & 7691 & 226 & 148 & 1.007 & 8.153 & 1 & \\ 
487 & 0.5111 & 201.0 & 1.2 & 5450 & 125 & 63 & 1.154 & 1.932 & 2 & B2084AB\\ 
487 & 2.993 & 130.2 & 5.1 & 5450 & 125 & 374 & 1.005 & 10.138 & 1 & \\ 
494 & 2.2257 & 153.7 & 5.7 & 4985 & 100 & 222 & 1.003 & 13.599 & 1 & \\ 
498 & 2.1845 & 251.5 & 3.5 & 6218 & 176 & 384 & 1.02 & 4.914 & 3 & RST4412AB\\ 
522 & 0.753 & 241.1 & 4.9 & 7381 & 109 & 82 & 1.005 & 9.342 & 1 & \\ 
527 & 0.2211 & 255.9 & 2.1 & 7403 & 163 & 36 & 1.07 & 2.738 & 1 & \\ 
550 & 0.6009 & 116.1 & 1.7 & 5076 & 62 & 37 & 1.1 & 2.269 & 1 & \\ 
563 & 0.3841 & 33.4 & 0.8 & 5383 & 157 & 60 & 1.216 & 1.694 & 1 & \\ 
568 & 0.2946 & 190.7 & 0.8 &  & 305 & 89 & 1.216 &  & 2 & HDS1310 \\ 
570 & 2.1638 & 110.6 & 6.6 & 5973 & 176 & 380 & 1.001 & 19.856 & 1 & \\ 
575 & 0.6293 & 13.9 & 4.7 & 6529 & 173 & 108 & 1.007 & 8.367 & 1 & \\ 
579 & 0.3303 & 202.6 & 3.3 & 4867 & 456 & 150 & 1.024 &  & 1 & \\ 
593 & 0.5794 & 169.7 & 3.9 & 7034 & 411 & 238 & 1.014 & 5.597 & 1 & \\ 
594 & 2.8688 & 45.3 & 0.4 &  & 308 & 883 & 1.301 &  & 3 & B1595\\ 
598 & 1.7109 & 290.5 & 6.2 & 8324 & 581 & 994 & 1.002 & 16.803 & 1 & \\ 
601 & 0.1512 & 352.5 & 3.1 & 6106 & 255 & 38 & 1.028 & 4.219 & 1 & \\ 
602 & 0.976 & 164.4 & 3.7 & 7649 & 615 & 600 & 1.016 & 5.436 & 1 & \\ 
605 & 0.6548 & 315.5 & 1.1 & 8265 & 508 & 332 & 1.168 & 1.938 & 2 & JSP296 \\ 
608 & 0.248 & 44.4 & 0.1 & 8323 & 500 & 124 & 1.383 & 1.397 & 2 & RST2587\\ 
609 & 0.3001 & 91.1 & 4.0 & 6272 & 154 & 46 & 1.012 & 6.107 & 1 & \\ 
611 & 2.3787 & 93.3 & 5.5 & 5689 & 96 & 228 & 1.003 & 12.33 & 3 & \\ 
612 & 2.0252 & 98.7 & 3.9 & 6422 & 284 & 575 & 1.014 & 5.69 & 1 & \\ 
612 & 0.1728 & 145.2 & 1.5 & 6422 & 284 & 49 & 1.119 & 2.079 & 1 & \\ 
619 & 0.2797 & 234.6 & 1.0 & 5096 & 806 & 225 & 1.182 &  & 1 & \\ 
621 & 2.0655 & 19.2 & 6.5 & 7850 & 186 & 384 & 1.001 & 19.256 & 1 & \\ 
630 & 0.2099 & 315.8 & 3.6 & 6849 & 295 & 61 & 1.018 & 5.245 & 1 & \\ 
635 & 1.7656 & 260.4 & 4.8 & 5914 & 58 & 102 & 1.006 & 9.176 & 1 & \\ 
637 & 2.3926 & 332.9 & 5.1 & 5637 & 63 & 150 & 1.005 & 10.059 & 3 & \\ 
640 & 0.2498 & 85.2 & 4.8 & 6587 & 341 & 85 & 1.006 & 8.756 & 1 & \\ 
642 & 0.9513 & 238.0 & 2.6 &  & 422 & 401 & 1.045 &  & 1 & \\ 
644 & 1.758 & 132.4 & 0.4 & 6112 & 1330 & 2338 & 1.301 &  & 2 & RST2447 \\ 
645 & 4.0745 & 109.0 & 1.9 & 5415 & 391 & 1593 & 1.083 & 2.505 & 3 & \\ 
649 & 2.6777 & 130.3 & 3.2 &  & 430 & 1151 & 1.026 &  & 1 & \\ 
651 & 3.608 & 191.7 & 2.2 &  & 86 & 310 & 1.064 &  & 3 & BU17AB \\ 
658 & 3.88 & 67.2 & 0.7 & 6521 & 245 & 950 & 1.235 & 1.627 & 3 & HJ4275\\ 
659 & 3.8774 & 67.2 & 1.0 & 5990 & 201 & 779 & 1.182 & 1.779 & 3 & \\ 
666 & 0.248 & 257.4 & 2.8 & 6680 & 149 & 36 & 1.037 & 3.601 & 1 & \\ 
676 & 1.5023 & 260.6 & 1.5 & 5430 & 545 & 818 & 1.119 & 2.151 & 3 & \\ 
680 & 0.7809 & 331.0 & 4.3 & 5967 & 158 & 123 & 1.009 & 6.942 & 1 & \\ 
684 & 0.2974 & 303.3 & 3.1 & 9488 & 580 & 172 & 1.028 & 4.103 & 1 & \\ 
690 & 0.2221 & 292.4 & 1.2 & 5538 & 144 & 31 & 1.154 & 2.026 & 1 & \\ 
697 & 1.1597 & 138.2 & 4.9 & 5447 & 92 & 106 & 1.005 & 9.255 & 1 & \\ 
697 & 0.0709 & 165.9 & 0.2 & 5447 & 92 & 6 & 1.353 & 1.43 & 1 & \\ 
703 & 1.4153 & 221.8 & 4.7 & 5384 & 111 & 157 & 1.007 & 8.449 & 1 & \\ 
758 & 0.1738 & 349.0 & 1.1 & 6072 & 154 & 26 & 1.168 & 1.907 & 1 & \\ 
759 & 2.694 & 234.6 & 3.6 & 6107 & 660 & 1778 & 1.018 & 5.257 & 3 & \\ 
772 & 2.3419 & 165.3 & 5.8 & 5184 & 130 & 304 & 1.002 & 14.246 & 3 & \\ 
779 & 1.5378 & 165.3 & 4.8 &  & 273 & 419 & 1.006 &  & 1 & \\ 
831 & 0.167 & 327.9 & 1.4 & 5808 & 86 & 14 & 1.129 & 1.941 & 2 & RST1830\\ 
832 & 3.2133 & 63.9 & 4.7 & 5623 & 278 & 893 & 1.007 & 8.384 & 1 & \\ 
837 & 2.3128 & 281.7 & 4.6 & 6513 & 136 & 314 & 1.007 & 7.995 & 3 & \\ 
847 & 1.2446 & 317.1 & 0.7 & 6055 & 659 & 820 & 1.235 &  & 3 & \\ 
851 & 1.8583 & 253.8 & 5.2 & 5782 & 155 & 288 & 1.004 &  & 3 & \\ 
905 & 2.2757 & 100.8 & 5.9 & 5565 & 153 & 348 & 1.002 & 15.332 & 3 & \\ 
906 & 1.246 & 50.8 & 2.8 & 5954 & 138 & 171 & 1.037 & 3.575 & 2 & RST805\\ 
907 & 3.6665 & 52.0 & 1.1 & 6272 & 307 & 1125 & 1.168 & 1.852 & 3 & \\ 
914 & 0.104 & 171.3 & 0.1 & 5321 & 233 & 24 & 1.383 & 1.35 & 1 & \\ 
926 & 0.1903 & 159.6 & 2.2 & 5621 & 216 & 41 & 1.064 & 2.802 & 1 & \\ 
930 & 0.692 & 32.6 & 2.5 & 6380 & 99 & 68 & 1.049 &  & 2 & B1455\\ 
931 & 0.1118 & 56.8 & 0.3 & 6434 & 691 & 77 & 1.326 &  & 1 & \\ 
952 & 1.1682 & 135.0 & 4.7 & 7025 & 461 & 538 & 1.007 & 8.607 & 1 & \\ 
952 & 0.1176 & 59.4 & 3.6 & 7025 & 461 & 54 & 1.018 & 5.245 & 1 & \\ 
954 & 2.3471 & 50.2 & 6.2 & 5820 & 232 & 544 & 1.002 & 15.699 & 1 & \\ 
1033 & 0.2684 & 220.4 & 1.7 & 6113 & 219 & 58 & 1.1 & 2.367 & 1 & \\ 
1049 & 1.2782 & 154.2 & 5.6 & 6599 & 389 & 497 & 1.003 & 12.617 & 1 &

\end{longtable*}
\textbf{Table 1 Notes. -- }Columns (1-4) gives the properties of companions to TOIs detected by SOAR. Uncertainties for these measurements and the observation epoch are provided in Table \ref{tab:whitelist}. Columns (5-6) gives the effective temperature and distance to the TOI given in the TIC \citep{tic}. Column (7) gives the projected separation of the companion (assuming it is physically associated with the primary), derived from the on-sky separation measured by SOAR and the distance to the star. Columns (8-9) give the radius correction factor for hosted planets in each system due to the contamination from the detected star in the scenario where the primary is the planetary host and the scenario in which the physically associated secondary is the planetary host. Column (10) is a flag denoting a previous detection of each companion. The flags are \textbf{(1)}: new pair, contamination not included in the TIC; \textbf{(2)}: known pair, contamination not included in the TIC; \textbf{(3)}: known pair, contamination included in the TIC. Column (11) provides the discoverer designation code if the companion is in the Washington Double Star Catalog maintained by the USNO. Explanations for codes are available at \url{https://www.usno.navy.mil/USNO/astrometry/optical-IR-prod/wds/WDS}.
\end{ThreePartTable}
\end{tiny}

\begin{figure*}
    \centering
    \includegraphics[scale=0.9]{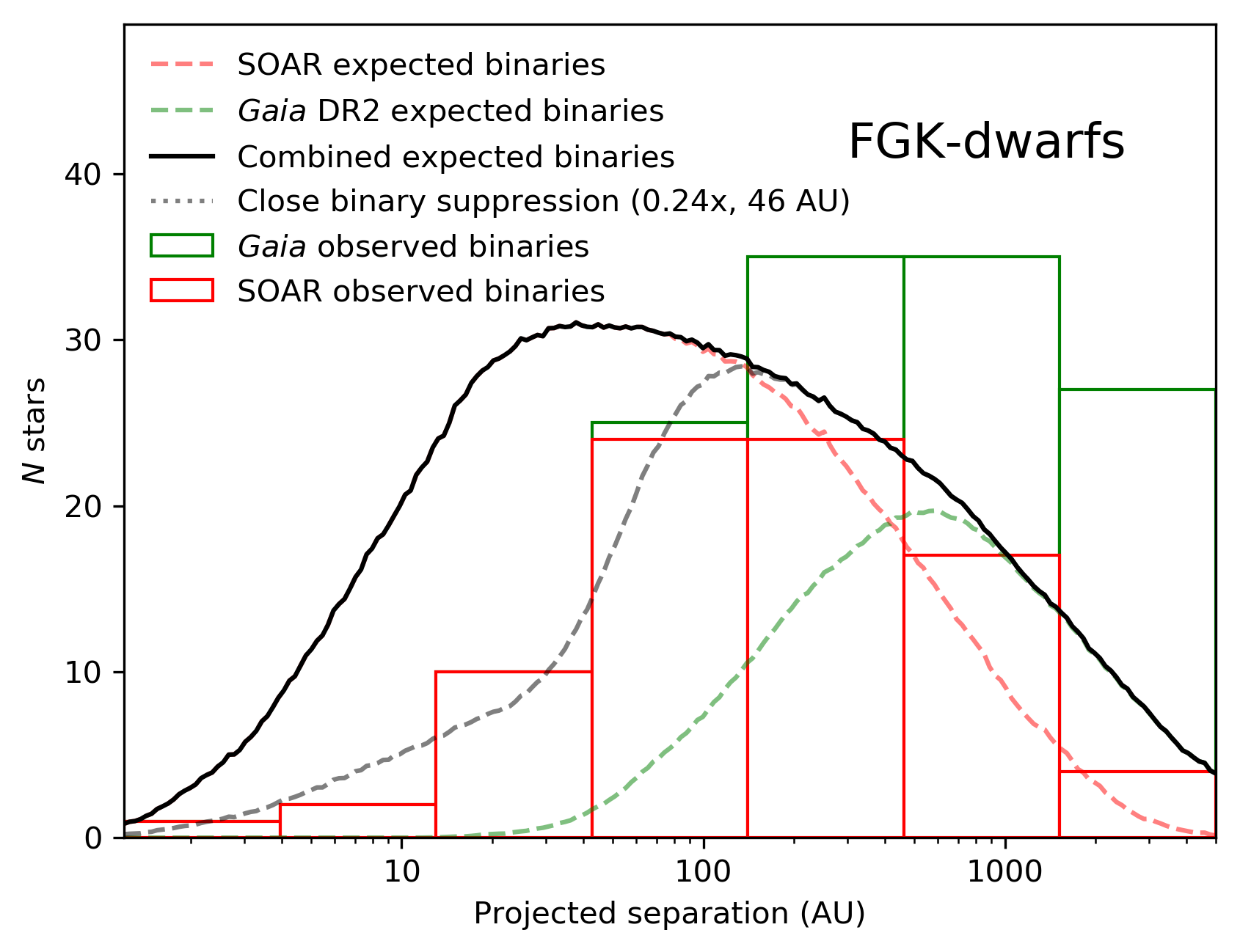}
    \caption{In red and green, the number of observed companions from SOAR and in Gaia DR2 for solar-type TESS planet candidate hosts in logarithmic bins of projected separation of 0.5 dex width. Companions found in both SOAR and Gaia are included in the SOAR sample. In black is the expected distribution from a multiplicity study of field stars \citep{raghavan10}, combining both field binaries that would be detected by SOAR and Gaia. The expected binaries from SOAR and Gaia, individually, are also plotted. These distributions take into account the detection sensitivity of both SOAR and Gaia. The observed distribution shows a clear paucity of TESS planet candidate host binaries at small projected separations compared to the field stars, and the inverse at wide separations. A best-fit model for the close binary suppression applied to the expected distribution of binaries for field stars is also plotted. The best-fit model has binaries suppressed by a factor of 0.24 at physical separations less than 46 AU.}
    \label{fig:tessvsfield}
\end{figure*}

\subsection{Implications for TESS planet radii}

As discussed in Section \ref{sec:radiuscorrections}, the additional flux from a stellar companion will dilute the transit signal in the TESS light curves, resulting in an underestimated planetary radius. We report the radius correction factors for planet candidates hosted in resolved binaries in our survey in Table \ref{tab:binaries}.

In general, the identity of the host star for an S-type planet in a close binary is ambiguous \citep{horch14}, although there is evidence that typically the primary is more likely to be the planet host \citep{gaidos16}. We, therefore, report correction factors for each host scenario. Overall, we find a mean correction factor of 1.11 in the cases where all the planets orbit the primary stars. This is similar to the factor of 1.08 found for \textit{Kepler} planets \citep{ziegler18a} under the same assumption. Likewise, if all planets orbit the secondary stars which are bound to the primary\footnote{Not every TESS planet candidate host had stellar radius estimates in the TIC or Gaia DR2. These targets are only included in calculating the mean correction factors in the case in which the primary star is assumed to be the host. In this case, only the flux contribution of the primary star is needed to determine the correction factor, as seen in Equation \ref{eq:1}.}, the radii of the planets will increase by a factor of 2.55 on average. This is slightly less than 3.29 found for \textit{Kepler} planets by \citet{ziegler18a}, which is likely due to the lower number of field stars detected in the TESS sample (see Section \ref{sec:unboundedstars}). Indeed, if a faint field companion is considered as bound, its estimated radius is small, and the resulting correction factor for the radius of a planet orbiting this companion becomes large.

In perhaps a more realistic scenario where planets are equally likely to orbit the primary or secondary star, we find an average correction factor of 1.82. This again is slightly lower than the correction factor of 2.18 found with similar assumptions in the \textit{Kepler} survey.

We can restrict our separation range to reduce the fraction of unassociated stars in our sample. Within 1$\arcsec$, we find mean correction factors of 1.14, 1.90, and 1.55 under the assumptions that all primary stars host the planets, all secondary stars host the planets, and that either star is equally likely to host the planet, respectively. The latter figure, more probable than either of the other cases, is in agreement with the radius corrections of 1.6, 1.64, and 1.54  found for the \textit{Kepler} planets by \citet{ciardi15}, \citet{hirsch17}, and \citet{ziegler18a}, respectively.

\subsection{Close binary inference with Gaia DR2}

While Gaia DR2 typically can not resolve binaries with separations less than approximately 0\farcs7 \citep{roboaogaia}, the additional source does often result in spurious astrometric solutions. The reliability of the Gaia astrometry is quantified by the re-normalized unit weight error (RUWE)\footnote{Described in detail in the Gaia DPAC public documentation, available at \url{https://gea.esac.esa.int/archive/documentation/GDR2/Gaia_archive/chap_datamodel/sec_dm_main_tables/ssec_dm_ruwe.html}}, which is near 1.0 for single sources, with a greater value (for instance, $>$1.4) indicating a non-single or otherwise extended source. Sources with only a two-parameter astrometric solution have null RUWE values.

We find that for the 135 observed TESS planet candidate hosts observed with SOAR with RUWE values $>$1.4, 114 had resolved companions. Twelve of the observed targets had null RUWE values and all 12 had bright, close companions ($\Delta$mag$<2$ and 0\farcs1$<\rho<0\farcs5$). For the observed targets with either high or null RUWE values, approximately 86$\%$ had companions, typically within the Gaia DR2 binary resolution limit of 0\farcs7. The median RUWE value for resolved close binaries ($\rho$$<$0\farcs75) is 5.56, compared to 1.04 for wider binaries ($\rho$$>$0\farcs75) and 1.03 for single targets. Only 6 targets with close binaries had RUWE values less than 1.4, five (TOIs 264, 476, 575, 609, and 640) of which have high contrasts ($\Delta$mag$>$4) and one (TOI-379) has a very low separation ($\rho$=0\farcs03). The RUWE values and properties of resolved systems are plotted in Figure \ref{fig:gaiaruwe}.

It is unclear why some single stars (21 out of 412 observed) have high RUWE values. One possibility is the number of Gaia observations of each star. Gaia uses a scanning law that passes through the north and south ecliptic poles every six hours, resulting in approximately twice as many observations at mid-ecliptic latitudes as near the ecliptic plane or poles \citep{gaia2016}. A Kolmogrov-Smirnov test, however, finds the distribution of ecliptic latitudes for single stars with high and low RUWE values to be similar. The distribution of Gaia magnitudes and nearby stellar densities (determined in Section \ref{sec:unboundedstars}) of single stars with high and low RUWE values were also similar. 

It is also possible that the anomalously high RUWE values for single stars are a result of binaries not detected by the SOAR speckle observations, such as those with close separations or high magnitude differences outside the sensitivity of SOAR. Further observations, with larger aperture telescopes and laser guidestar adaptive optics, could confirm whether these single stars have companions. Of the 21 single stars with high RUWE values, five have the results of additional high-resolution imaging available on ExoFOP.\footnote{TOI-271 was observed with NaCo on the VLT (posted by Elisabeth Matthews), TOI-502 was observed with NIRI on Gemini-North (Elisabeth Matthews) and NIRC2 on Keck-2 (David Ciardi), TOI-674 was observed by NIRI on Gemini-North (Ian Crossfield/Elisabeth Matthews), and TOI-254 and 311 was observed by PHARO on Palomar 5-m (David Ciardi).} None of the five targets had a detected companion star.

Checking for a high or null RUWE value can serve as an excellent first check for potential companion stars in TESS systems, although further high-resolution observations would still be required to determine the properties of a purported companion. The clustering of binary systems with null RUWE values in a region of similar separation and magnitude difference may, however, could even be used to infer the vague properties of a subset of the unresolved stars in Gaia DR2.

\section{Impact of Binary Stars on the TESS Planetary Systems}\label{sec:analysis}

The presence of a binary companion can result in a dynamically harsh environment, reducing the probability that a planetary system can form and survive. Some planets are found in close binary systems, however. In this section, we search for further insight into the impact binary stars have on the TESS planet population.

\subsection{Preparation of the sample}
\label{sec:preparation}

To facilitate an analysis into the multiplicity of our observed targets, some sample preparation was required. 


The speckle imaging is a snapshot of the host systems, providing an on-sky angular separation between the primary and secondary star. To determine the projected physical separations, the distance to each system is required. We collect distances to each of these targets from Gaia DR2 \citep{bailerjones18}, when available. However, Gaia can provide spurious astrometric solutions in the case of close, unresolved binaries \citep{arenou18}. So, when Gaia distance errors are large (greater than 20\%), we use the distances reported in the TIC \citep{tic}, which were derived using inverse Stefan-Boltzmann relations based on $V$ magnitudes. The distances used for each target in this analysis are available in Table \ref{tab:binaries}. 

We prepare our sample by removing stars with T$_{eff}$ in the TIC inconsistent with an FGK-type star (i.e., T$_{eff}>7200$ K and $<3900$ K), using the relations of \citet{pecaut12}. We also remove binaries with contrasts indicating mass ratios $q<0.4$. These systems, with high magnitude differences, are significantly more likely to be chance alignments based on the analysis in Section \ref{sec:unboundedstars}, and their exclusion enables comparisons between the \textit{Kepler} \citep{kraus16} and TESS sample. We determine $q$ by finding the mass of the primary star based on its likely spectral type, estimated using the T$_{eff}$ reported in the TIC, and the secondary star based on the magnitudes difference with the primary \citep{kraus07}. We also remove systems with a TESS follow-up disposition of false positive, and systems with only a single transit detected by TESS. After these cuts, our sample includes 455 stars observed with SOAR.

To improve our coverage of wide binaries, we include companions to these 455 SOAR targets found in Gaia DR2 \citep{gaia} with proper motions and distance estimates \citep{bailerjones18} consistent at 2$\sigma$. We search out to an on-sky angular radius equivalent to a 5000 AU projected separation based on the Gaia distance estimate. \citet{elbedry18} found that the proper motion of wide binaries can vary significantly because of orbital motion. For each star, we calculate the maximum Keplerian orbital on-sky motion as a function of projected separation. The proper motion of each nearby Gaia star is allowed to vary by this orbital motion in our binary detection. The Gaia DR2 binaries used are listed in Table \ref{tab:gaiabinaries} in the Appendix. Only pairs of stars with contrasts consistent with mass ratios $q>0.4$ were included. \citet{roboaogaia} provided the binary recovery sensitivity of Gaia DR2.

\subsection{Multiplicity of solar-type TESS planet candidate hosts}\label{sec:multiplicity}

Close binaries can provide many potential obstacles to planet formation and evolution. In a large survey of \textit{Kepler} planets, \citet{kraus16} found that far fewer planets were detected around stars with companions at solar system scales, within approximately 50 AU. The TESS sample is quite disparate in several ways from the \textit{Kepler} sample, as shown in Figure \ref{fig:histogram}. In general, the TESS planets are somewhat larger and at shorter periods than the \textit{Kepler} planets, a consequence of the TESS photometric precision and survey strategy. Unlike \textit{Kepler}, the TESS planets are spread across the sky and sample a more diverse set of the Galactic stellar population, providing an opportunity to confirm and characterize the effect of binaries

To understand how binaries impact planetary systems, we compare our sample to a simulated survey of field solar-type stars. We use the field binary statistics of \citet{raghavan10}, who found a flat eccentricity distribution, a log-normal period distribution (with a mean of log $P$ = 5.03, corresponding to an orbital semi-major axis of approximately 50 AU, and $\sigma_{log P}$ = 2.28), and a nearly uniform mass ratio distribution (with a sharp increase near equal mass ratio) in the population of solar-type field binaries. We follow the procedures of \citet{kraus16} to account for projection effects, Malmquist bias, and the detection limits of our survey. We also account for the reduced sensitivity of TESS to planet transits due to dilution by the stellar companion.

For each solar-type star observed in our survey, a Monte Carlo model was constructed to determine the expected number of binary companions at a range of projected separations between 1-5000 AU. In each of 10$^{5}$ iterations, there was a 33\%$\pm$2\%, 8\%$\pm$1\%, and 3\%$\pm$1\% probability that one, two, or three companion stars would be populated, respectively (the observed multiplicity of solar-type stars) . Since binaries are over-represented in flux-limited surveys \citep{schmidt68}, we correct for Malmquist bias by adjusting this probability by an additional factor equal to the fractional volume excess in binaries due to their relative brightness, V$_{bin}$/V$_{single}$. The period, eccentricity, and mass ratio of these binaries were drawn from the distributions reported in \citet{raghavan10}. The period was converted to a semi-major axis using the TIC estimated stellar masses. We select uniformly distributed values for the cosine of inclination, the position angle of the ascending node, the longitude of periastron, and the time of periastron passage. Finally, the instantaneous separation was projected to the distance to the primary star as reported in Gaia DR2. The mass ratio was converted to an approximate magnitude contrast using the relations in \citet{kraus07}, and possible detection by SOAR speckle imaging and Gaia DR2 was determined using the measured sensitivity limits and the companion's contrast and separation. We use the ratio of non-detected binaries to the total number of binaries at each separation to determine a completeness correction due to limitations in the ability to resolve close or wide companions.

\begin{figure}
    \centering
    \includegraphics[scale=0.65]{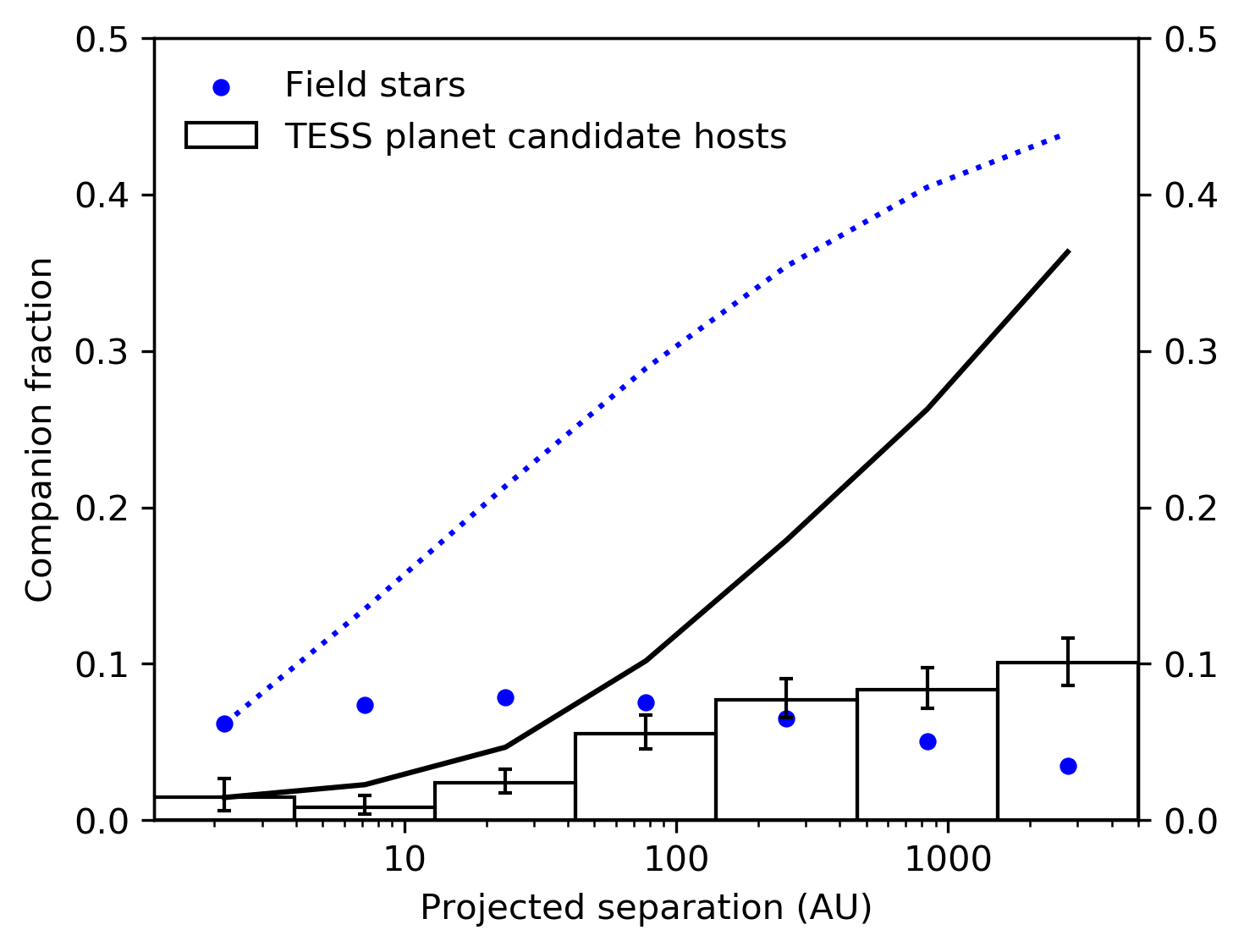}
    \caption{The observed companion rates for the solar-type TESS planet candidate hosts, corrected for survey completeness, for logarithmic bins with a width of 0.5 dex is shown by the black bars. The completeness factors are  based on the binary statistics of \citet{raghavan10} and our detection limits. The estimated companion rates for field stars, plotted by blue points, are included for comparison. The cumulative distributions for the TESS planet candidates and field stars are plotted as dotted and solid lines, respectively.}
    \label{fig:companionrates}
\end{figure}

The resulting observed binaries of TESS planet candidate hosts from SOAR and Gaia compared to the expected number derived for field stars are shown in Figure \ref{fig:tessvsfield}. The uncertainty in the expected number of observed binaries at each separation range is derived from the spread of binaries in the simulated surveys, which propagates the field binary rate uncertainties reported by \citet{raghavan10}. The observed companion rate to the TESS planet candidates as a function of projected separation was determined by dividing the number of observed binaries by the total number of stars observed. The companion rate in each separation bin was then corrected for survey completeness using our measured sensitivity limits. The companion rates for the TESS planet candidate hosts and field solar-type stars are shown in Figure \ref{fig:companionrates}. The distributions differ substantially, both at close and wide separations, which we will address in turn.

\subsubsection{Bias against planet detection in binaries}

The additional flux from a companion star reduces the depth of the transit, making planet detection more difficult, and the TIC \citep{tic} was constructed without consideration for potential stellar multiplicity. Left uncorrected, this observational bias would result in fewer binaries being detected among planet hosts, as transiting planets are easier to detect around single stars with no flux dilution.

We quantify the magnitude of this bias using the methods of \citet{wang15}, adapted for the TESS sample. The planet search pipeline for TESS, run by the Science Processing Operations Center \citep[SPOC,][]{jenkins16}, flags a potential transit candidate if the signal-to-noise ratio (S/N) is greater than 7.1. The S/N can be calculated by the equation:
\begin{equation}
    S/N = \frac{\delta}{CDPP_{eff}}\sqrt{N_{transits}},
\end{equation}
where $\delta$ is the transit depth, CDPP$_{eff}$ is the effective combined differential photometric precision per transit, and $N_{transits}$ is the number of observed transits. In the presence of a companion star, the transit depth is reduced due to the additional flux by a factor $X$ given in Equations \ref{eq:1} and \ref{eq:2}, depending on the identity of the host star, which likewise reduces the S/N of the detection.

We estimate the detection bias using a simulation for each of the TESS systems observed with no binary companion. For every single system, we choose the planet that has the highest S/N and add a companion star, whose mass is drawn randomly from the binary mass distribution described in \citet{raghavan10} (e.g., uniform with a spike near unity). The transit depth of the planet is calculated for one of two randomly selected scenarios: 1) the planet orbits the primary star, with reduced transit depth from the second star, 2) the planet orbits the secondary with updated transit depth due to the different stellar radius and dilution from the primary star. For the selected scenario, the S/N of the detection is calculated using the new transit depth and, if greater than 7.1, it was determined the planet could still be detected by TESS with the companion star. This procedure is repeated 10$^{5}$ times for each observed TESS system to determine $\alpha$, the probability of a planet detection in that system if a stellar companion is present. The distribution of $\alpha$ for the observed systems is an indication of the magnitude of the binary detection bias (i.e., if most of the observed planets in single systems would still be detectable even with a binary companion, then it is likely few additional planet signals were not detected by TESS due to dilution from a second star). Relevant data and $\alpha$-values for each system are provided in the Appendix in Table \ref{tab:binarybias}.

We find that most of the TESS systems would still be detectable even with an equal mass companion, with median and mean $\alpha$ values of 1.0 and 0.97.  Significantly more TESS planets are detectable with companions compared to the \textit{Kepler} sample of \citet{wang15}, who found a median value of $\alpha$=0.89. This is likely due to the larger radii and lower periods of the TESS planets, which result in higher S/N. We would expect 9$\pm$2 of the 416 observed single systems (2.1$\pm$0.4\%) would not have planets detected by TESS if each were in stellar binaries. Therefore, the number of potential TESS planets that were not detected due to binary dilution is likely to be small relative to the number of systems observed, and we expect the detection bias against planets in binary systems does not significantly impact the subsequent analysis.

In the simulations, the stellar radii are from the TIC. The stellar radii of secondary stars were determined using the mass-radius relationship of \citet{feiden12}. The CDPP$_{eff}$ was estimated from the TESS band magnitude of the primary and the photometric precision model of \citet{stassun18}, interpolated based on the transit duration.

\subsubsection{Suppression of planet-hosting close binaries}
We find significantly fewer binaries in the TESS sample with projected separations of less than 100 AU than would be expected for a similar survey of field stars (44 observed binaries compared to an expectation of 124$\pm$8\footnote{Uncertainties in the expected number of observed binaries are derived from the resulting distribution of simulated surveys, as described in Section \ref{sec:multiplicity}.}, a 9.0$\sigma$ discrepancy). We find a completeness corrected companion rate for TESS planets hosts of 9.8$^{+1.5}_{-1.3}\%$ at projected separations of less than 100 AU and larger than 1 AU. For comparison, we estimate for field stars a companion rate at similar projected separations of 27.7$^{+2.0}_{-2.2}\%$, using the binary statistics of \citet{raghavan10}. TESS planet hosts are therefore approximately 3$\times$ less likely to have a close binary companion, compared to field stars. 

\begin{figure}
    \centering
    \includegraphics[scale=0.6]{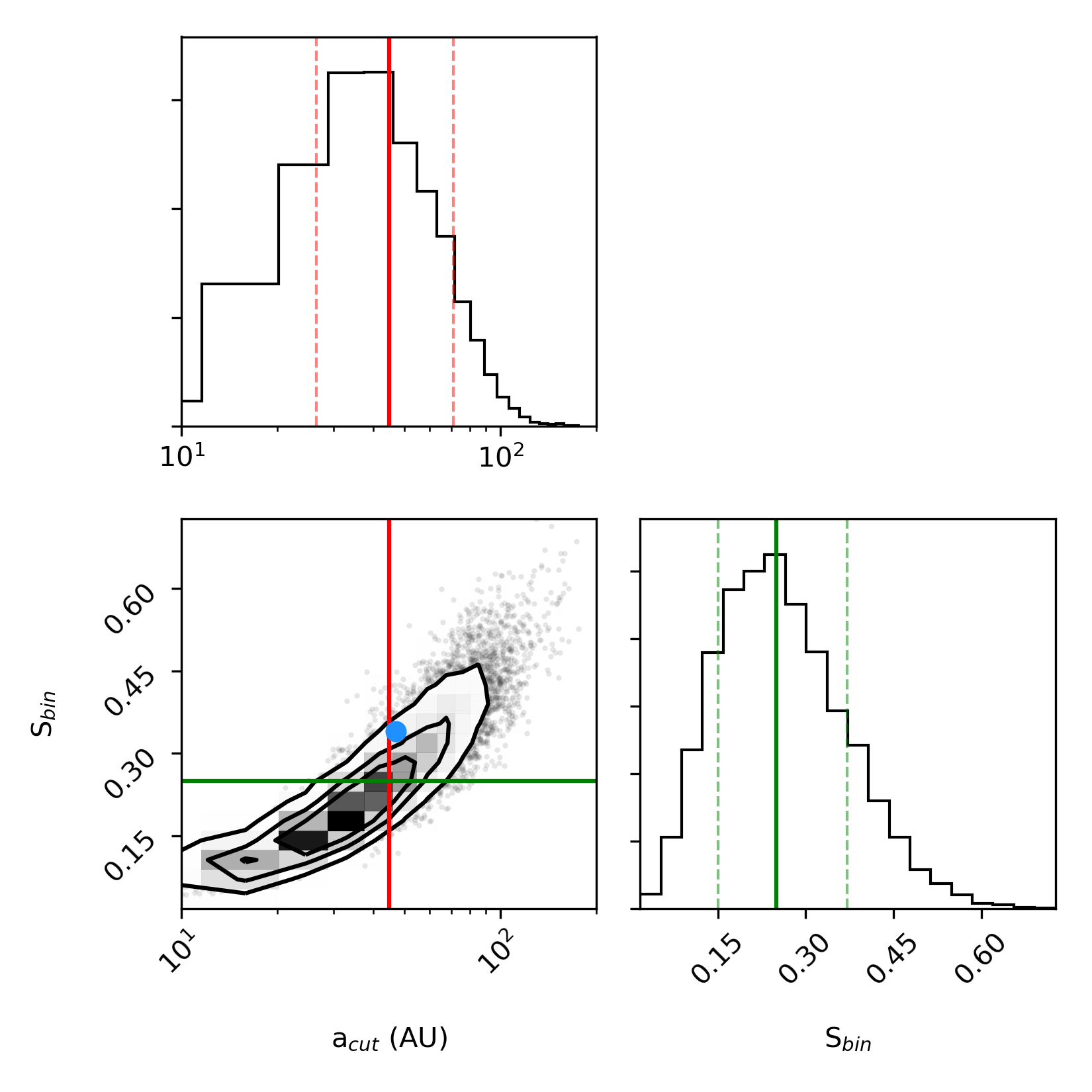}
    \caption{The distributions of suppression factors and semi-major axis cuts from 10$^{6}$ chains of an MCMC analysis to model the observed close binary suppression seen in systems with TESS planet candidates. Solid lines indicate the median value of each distribution (a$_{cut}$=46 AU and S$_{bin}$=0.24) and dashed lines mark the 68$\%$ confidence interval. The two parameters are correlated, such that less suppression is required if the semi-major axis cutoff is larger, and vice versa. A blue circle marks the values of the best fit model from the \textit{Kepler} sample of \citet{kraus16}.}
    \label{fig:suppressioncorner}
\end{figure}

Similar to \citet{kraus16}, the dearth of close binaries for planet hosts can be modeled by a simple two parameter model, using a suppression factor, $S_{bin}$, that cuts on at some semi-major axis value, $a_{cut}$. We performed a Markov chain Monte Carlo (MCMC) analysis to explore 10$^{6}$ possible values for $S_{bin}$ and $a_{cut}$, seeking to reduce the $\chi^{2}$ goodness of fit to the observed distribution. The resulting distributions are shown in Figure \ref{fig:suppressioncorner}. We find an optimal values for the suppression with 68$\%$ credibility ranges to be $S_{bin}$=24$^{+11}_{-7}\%$ and $a_{cut}$=46$^{+22}_{-15}$ AU, shown in Figure \ref{fig:tessvsfield}. These values are in agreement with those found by \citet{kraus16} for \textit{Kepler} planet candidates ($S_{bin}$=34$^{+15}_{-14}\%$ and $a_{cut}$=47$^{+59}_{-23}$ AU), and again the null hypothesis (i.e., the field and planet candidate host distribution are similar with no binary suppression occurring) is strongly disfavored at 9.8$\sigma$.

\subsubsection{Enhancement of wide binaries in systems with large planets}\label{sec:widebinaries}

At wide separations, more binaries were detected around the TESS planet candidate hosts compared to the field stars: 119 observed binaries with projected separations greater than 100 AU, compared to an expected number of 77$\pm$7, a 4.9$\sigma$ discrepancy. 

\begin{figure}
    \centering
    \includegraphics[scale=0.52]{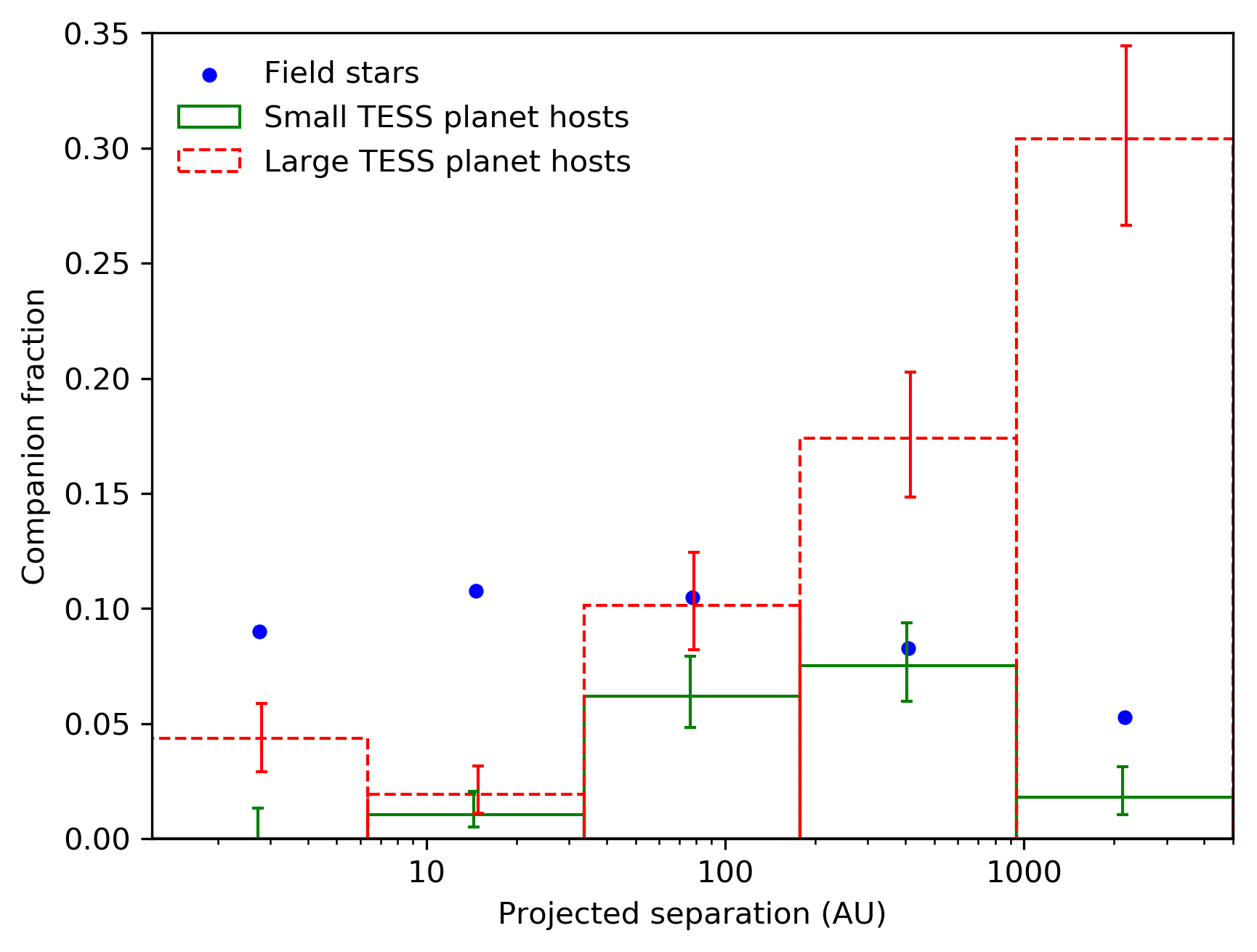}
    \caption{The completeness corrected companion fraction per 0.5 dex bins in projected separation for small and large TESS planet candidate hosts observed in this survey. For reference, the separation distribution of field binaries from \citet{raghavan10} is included. Both populations of TESS planet hosts have suppressed rates of close binaries, but have diverging binary rates at wide projected separations. Large planets (R$_{p} >$9 R$_{\oplus}$) are approximately 3.5$\times$ more likely to be found in a wide binary, compared to field stars. Conversely, small planets are 2$\times$ less likely to be found in a wide binary.}
    \label{fig:planetradius}
\end{figure}

The wide binaries being detected are almost exclusively those hosting the large planet candidates. This is readily apparent if we split our sample into two bins using a radius cut of 9 R$_{\oplus}$ (approximately the size of Saturn), as shown in Figure \ref{fig:planetradius}. We find that both populations of 244 small and 199 large planets exhibit a paucity of systems in close binaries. At wide separations, however, the companion rates for the two populations diverge: at projected separations greater than 200 and 10$^{3}$ AU, large planet hosts have a completeness corrected companion rate of 47.8$\pm3.8\%$ and 30.4$\pm3.4\%$, respectively. For small planet hosts, the companion rates for similar projected separation ranges are 6.4$^{+1.5}_{-1.1}\%$ and 1.8$^{+1.3}_{-0.7}\%$, respectively. Thus, the two populations have discrepant companion rates for projected separations greater than 200 and 10$^{3}$ AU with significances of 9.8$\sigma$ and 8.1$\sigma$, respectively.

The companion rates for the small and large planets at wide separations are also both significantly divergent from that of field stars. We estimate from the distribution of \citet{raghavan10} that field stars have a companion rate of 13.7$\%$ and 5.3$\%$ at projected separations greater than 200 and 10$^{3}$ AU, respectively. Therefore, large TESS planets are approximately a factor of 3.5 more likely to be hosted in a wide binary than would be expected. Conversely, small TESS planets are 2$\times$ \textit{less} likely to be found in a wide binary system than would be expected from field star statistics.

\subsection{Binary fraction of M-dwarf planet candidate hosts}

We observed 44 planet candidate hosts with T$_{eff}$ estimates in the TIC consistent with an M-dwarf (T$_{eff}<$3900 K, \citet{pecaut12}). We detected companions to 16 of these hosts, for a multiplicity fraction of 36$\pm$9\%. This is consistent with the field star M-dwarf multiplicity fraction of 26.8$\pm$1.4\% found by \citet{winters19}.

\begin{figure}
    \centering
    \includegraphics[scale=0.6]{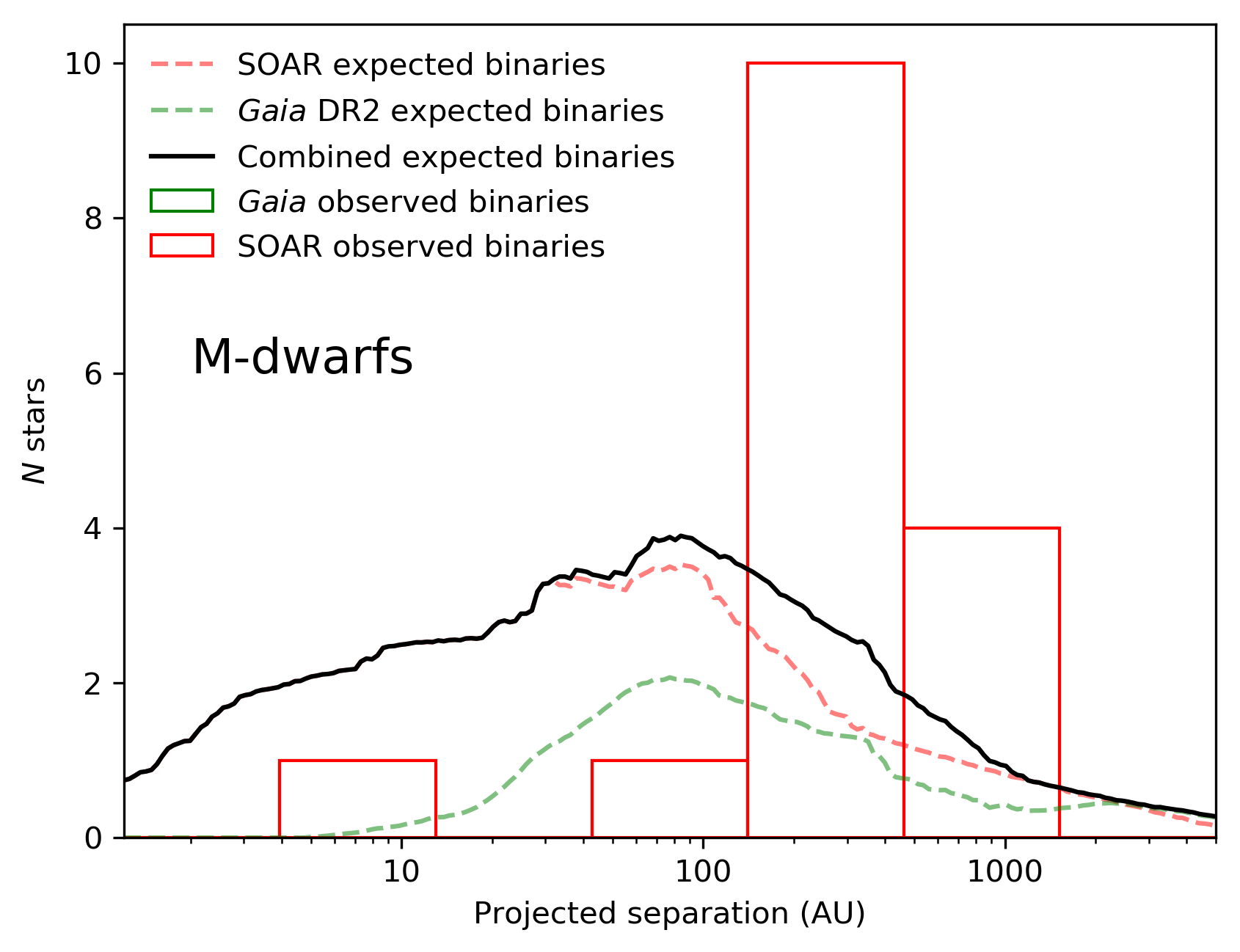}
    \caption{Similar to Figure \ref{fig:tessvsfield} for TESS M-dwarf planet candidate hosts, using the field binary statistics of \citet{winters19}. As with the solar-type stars, the M-dwarf planet candidate hosts have a deficit of close binaries at separations less than 100 AU, and a surplus of wide binaries separated by more than 100 AU.}
    \label{fig:mdwarfs}
\end{figure}

To compare the separation distribution of planet candidate M-dwarf hosts to the field star population we use the companion fraction, the log-normal projected separation distribution (peaking at 20 AU with $\sigma_{loga}$=1.16), and the uniform mass ratio distribution (with a slight increase in near-equal mass binaries) found by \citet{winters19}. The resulting distribution is shown in Figure \ref{fig:mdwarfs}.

We find fewer close binaries than would be expected to the M-dwarf planet candidate hosts: 2 observed compared to approximately 11 expected at projected separations of less than 100 AU. We also find a large number of companions at wider separations: 14 observed companions at projected separations between 100 and 5000 AU, compared to approximately 5 that would be expected.

Both of these results mirror those found with the solar-type sample. We do not see a large number of companions at very wide projected separations (s$>$1000 AU). This may, in part, be due to the M-dwarf projected separation distribution peaking at lower separations ($\sim$20 AU rather than $\sim$50 AU) and with lower variance than the solar-type sample. Also, as shown in Section \ref{sec:widebinaries}, the widest binaries typically host Jupiter-size planets, which are inherently rare around M-dwarfs \citep{dressing13}.

\subsection{Mass ratios of planet candidate host binaries}

The mass ratio, or $q$, distribution of solar-type binary systems was found to be nearly uniform by \citet{raghavan10}, with an increase for near-equal mass binaries. \citet{winters19} found a similar distribution at high-$q$ for M-dwarfs. \citet{ngo16} found that mass ratio distribution for hosts of hot Jupiters was heavily weighted towards low-$q$ companions.

The mass ratio distribution of resolved TESS planet candidate hosts may vary significantly from that of field stars. To compare the two populations, we use the mass ratios derived from the observed magnitude difference, as described in Section \ref{sec:preparation}, for the SOAR binaries and for the field stars, the mass ratios from the simulation described in Section \ref{sec:multiplicity}, which takes into account our survey sensitivity.

\begin{figure}
    \centering
    \includegraphics[scale=0.55]{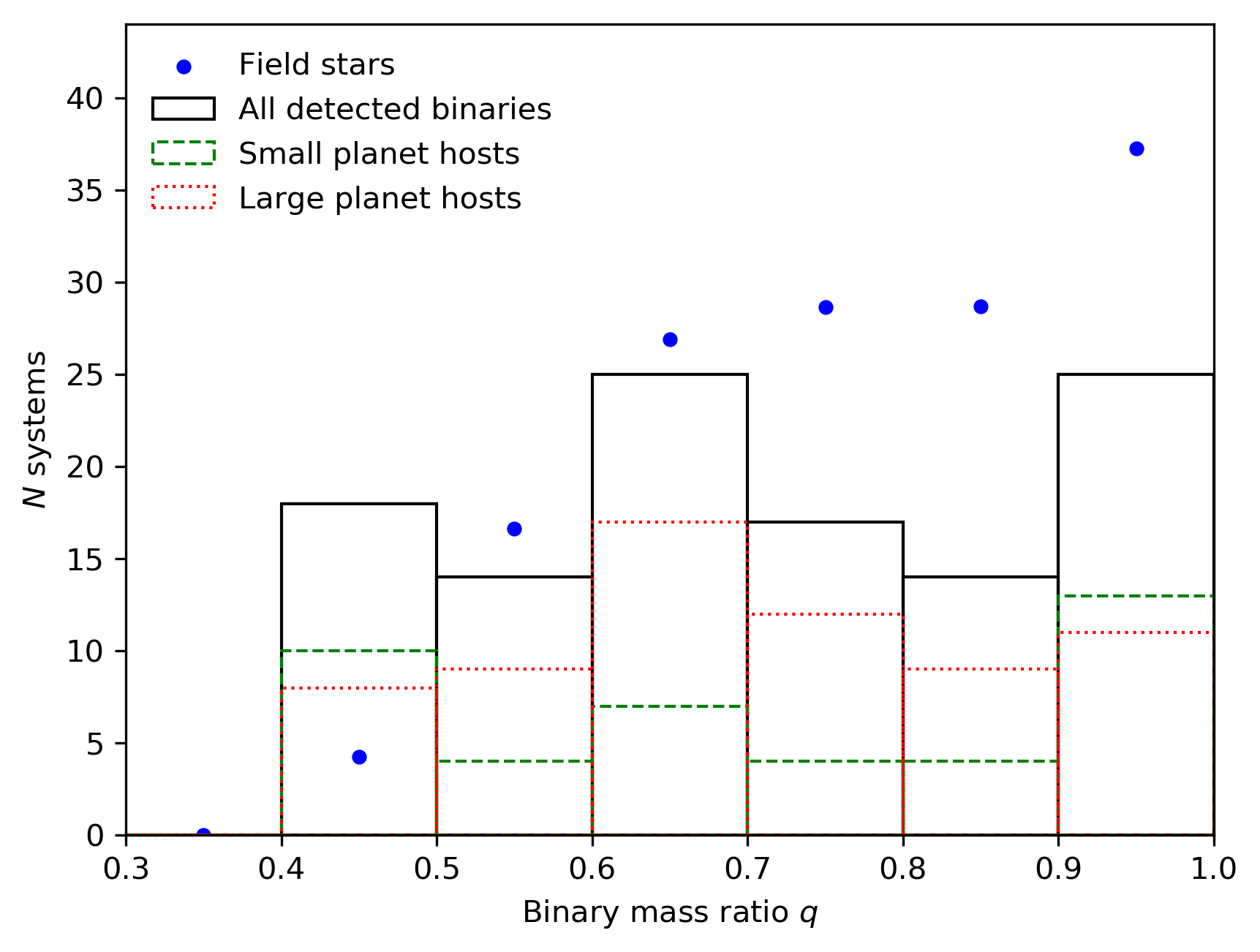}
    \caption{The mass ratio distribution of all observed binaries to TESS planet candidates resolved in SOAR speckle imaging, including the individual distributions for large and small planet candidate hosting stars. The expected distribution of observed binaries based on the near-uniform mass ratio distribution of field stars \citep{raghavan10} and the survey sensitivity is included. The observed mass distribution is consistent with uniform.}
    \label{fig:qs}
\end{figure}

The observed and expected mass ratio distribution of binaries to TESS planet candidate hosts is shown in Figure \ref{fig:qs}. The mass ratio distribution of observed binaries is consistent with being uniform for $q>0.4$; for lower mass ratios the SOAR sensitivity is low. There is no significant difference in the distribution for small and large planets (cut at 9 R$_{\oplus}$). Compared to the expected number based on field star statistics, we find slightly fewer high-$q$ binaries, due in part to binary suppression at low separations, and slightly more low-$q$ binaries, which is likely to be, at least in part, a consequence of unassociated field star contamination as the companions with large magnitude differences are at wide separations.

\section{Binary Impact on the Galactic Planet Population}\label{sec:discussion}

Approximately half of solar-type field stars in our galaxy are found with a stellar companion, and, as discussed in Section \ref{sec:intro}, the impact binaries have on the planet population is potentially large. The TESS sample is not statistically complete, containing a combination of small planets, similar to the majority of planets detected by \textit{Kepler}, and many large planets, similar to the extensively studied population of hot Jupiters \citep[e.g.,][]{wu07, knutson13, ngo16, evans16}. In our data, we see a combination of two separate effects found individually in previous studies: suppression of planets in binaries at low separations, and enhancement of binaries at wide separations.

Follow-up observations of the planet candidate hosts found in the \textit{Kepler} survey gave some insight into how binaries impact planetary systems. \citet{wang15} found some evidence in RV trends that binaries at separations $<$100 AU may suppress planet occurrence. \citet{kraus16} found that approximately a fifth of solar-type stars are not able to host planets due to the influence of stellar companions. Significant suppression in the rate of spectroscopic binaries hosting planets was detected by \citet{ngo16}, which helps rule out an unresolved population of very close stellar companions. We find similar suppression of close binaries among the TESS planet candidates, suggesting this effect is likely prevalent throughout the Galaxy, in regions of varying stellar density and age. The suppression of close binaries is apparent regardless of cuts in the orbital period, planetary radii, or stellar effective temperature. 

As noted by \citet{kraus16}, it is not clear why some close binaries are able to host planets with all of the theoretical obstacles to their formation and survival. Many of the TESS binaries that are close in projection likely have short orbital periods, meaning their motion may be detected over the coming years. Continued monitoring can provide orbital solutions for these systems to find a true physical separation rather than the projected separation presented here. Studying these orbits may provide insight into the conditions that exist such that planets may form and survive in the chaotic regime around close binaries.

It is possible that the enhanced companion fraction for small planets may be, at least in part, due to observation effects: the false positive rate of giant planets in transiting planet surveys has been found to be larger than for smaller planets \citep{fressin13}. The radii of Jupiter-mass planets and brown dwarfs are similar, and mass constraints are required to confirm each planet. However, brown dwarfs on close transits seem to be inherently rarer than massive planets \citep{bowler16}. In addition, while a planet is more likely to be hosted by the primary star in the majority of systems \citep{gaidos16}, multiple star systems will have some enhancement in planet occurrence due to having additional potential hosts.

The exclusion of previously confirmed planets from our target list, as discussed in Section \ref{sec:observations}, may result in a bias in our sample. Many of these confirmed planets were detected in ground-based surveys, which may avoid resolved binaries \citep{wasp, hats} or avoid following-up systems with binary indicators, such as multiple sets of spectral lines \citep{triaud17}. In addition, the contamination from unresolved, near-equal-mass binaries may result in the non-detection of planets by ground-based surveys \citep{bouma18}. Some of these planets may subsequently be detected by TESS. These observational biases could result in an inflated companion rate for the newly detected TESS systems in our survey.

Previous binary surveys of large planet hosts that were detected exclusively from ground-based surveys find a similar enhancement in binaries at wide separations. \citet{ngo16} found that stars hosting hot Jupiters were approximately three times more likely to have stellar companions than field stars. \citet{fontanive19} found a wide binary fraction (20-10,000 AU separations) for gas giants approximately twice that of field stars. Similarly, \citet{ziegler18b} found that the large, close-in \textit{Kepler} planets were significantly more likely to have companions than other populations of planets. We find a similar effect for our sample as a whole. However, closer analysis reveals this enhancement is due only to the systems hosting the largest planets. Indeed, a suppression effect is also seen for small planets in binaries at very wide projected separations ($s>$1000 AU).

There are two possible physical scenarios that could lead to a high companion fraction for systems hosting hot Jupiters: first, the binary companion may encourage in some manner the formation of large planets; and second, large planets form at similar rates in single and multiple star systems, but in binaries, the companion star drives large planets inward to the low-period regime probed by TESS. For the former, there is evidence that a stellar companion can lead to density waves in the protoplanetary disks \citep{dong15}. These high-density regions can subsequently seed the formation of planetesimals \citep{carrera15}. In addition, the protoplanetary disks around binary stars may have more mass than around single stars, which simulations suggest leads to larger planets \citep{mordasini12}.

The high number of gas giants in binaries coupled with the low number of observed smaller planets may be explained by planet-planet scattering during migration. In other words, the orbits of inner smaller planets may be altered by the gas giants being driven inward to low-period orbits. In one possible scenario, Kozai-Lidov instabilities induced by the stellar companion may initially drive the gas giant to a highly eccentric orbit \citep{holman97, naoz11}. \citet{ngo16} and \citet{fontanive19} found that the Kozai-Lidov effect alone was insufficient to explain the observed population of hot Jupiters. In scattering events, large planets on wide orbits are preferred, as quantified by the Safronov number \citep{ford08}. The highly eccentric gas giant in this scenario would dominate the inner planets as it nears perihelion, resulting in planet scattering events \citep{fabrycky07}. Eventually, the gas giant orbit will circularize to a low-period orbit due to planet-star tidal friction \citep{jackson09}. Observational evidence suggests planetary interactions during secular migration are not unusual: around a quarter of hot Jupiters are found on retrograde orbits, only possible through close planetary perturbations \citep{naoz11, ngo15}.

Dynamical interactions between planets with high mass disparities may dramatically alter the orbital inclination of the smaller planet, in many cases to a non-transiting orientation \citep{hamers17}, or drive the smaller planets to highly eccentric orbits \citep{xie16},  and possibly even ejection from the system \citep{davies14}. Planet-planet scattering has been shown to largely reproduce the observed distribution of eccentricities in transiting planets \citep[e.g.,][]{ford08, juric08, raymond11}. Numerical investigations suggest instabilities in giant planet orbits are likely destructive to inner terrestrial planets \citep{veras06, matsumura13}. Indeed, \textit{Kepler} found only a small fraction of small, close-in planets had gas giants in nearby orbits \citep{lissauer11, ciardi13, huang16}. Lastly, \citet{wang15} found that systems hosting small planets had fewer companions than field stars at separations up to 1500 AU, compared to 100 AU for systems with hot Jupiters. As they note, another possible explanation for this disparity is that the relative timescales of pericenter and nodal precession increases as planetary-mass decreases \citep{takeda08}. Thus for small planets, the Kozai timescale will be shorter than precession. Consequently, the weaker planet-planet coupling means smaller planets are more prone to the influence of distant stellar companions.

\section{Conclusions}\label{sec:conclusions}

We searched 542 TESS planet candidate hosts for companions using SOAR speckle imaging. We found 123 companions within 3$\arcsec$ of 117 targets. Contamination from these companions in the TESS light curves results in the radii of planet candidates in these systems to increase by a factor of 1.11, assuming the primary star is indeed the host. We find that TESS planet candidate hosts are around 3.5$\times$ less likely to have stellar companions at projected separations less than approximately 50 AU than field stars. The destructive impact of close binaries, previously seen in the \textit{Kepler} sample, is apparent in the local Galaxy. We also detect far more large planets, and far fewer small planets, in wide binaries then would be expected for field stars. This may be evidence of chaotic secular migration of gas giants, resulting from perturbations from the binary companion, inducing planet-planet scattering. The M-dwarfs hosting planet candidates have a similar binary pattern as the solar-type sample. The mass ratio distribution of planet candidate hosting stars is consistent with uniform, as is seen in field stars.

Future multi-band speckle observations by SOAR of the resolved binary systems hosting TESS planets will be able to determine the probability that the companions are indeed physically associated. In addition, multi-epoch observations over the coming years will be able to check for common proper motion and solve the orbits of bound systems, providing the semi-major axis and eccentricity of the binary systems hosting planets. Analysis of these binaries may provide insight into how some close systems were able to form and maintain their planetary populations. Lastly, the detection of Northern planet candidates by TESS, beginning in 2019, will provide many more nearby planet-hosting systems. Their proximity will allow efficient instruments on moderate size telescopes in the North, such as Robo-AO on the University of Hawaii 88-in \citep{automatedao}, to detect companions at Solar system scales.

\section*{Acknowledgments}

We thank the anonymous referee for her or his careful analysis and useful comments on the manuscript.

C.Z. is supported by a Dunlap Fellowship at the Dunlap Institute for Astronomy \& Astrophysics, funded through an endowment established by the Dunlap family and the University of Toronto. A.W.M was supported by NASA grant 80NSSC19K0097 to the University of North Carolina at Chapel Hill.

Based on observations obtained at the Southern Astrophysical Research (SOAR) telescope, which is a joint project of the Minist\'{e}rio da Ci\^{e}ncia, Tecnologia, Inova\c{c}\~{o}es e Comunica\c{c}\~{o}es (MCTIC) do Brasil, the U.S. National Optical Astronomy Observatory (NOAO), the University of North Carolina at Chapel Hill (UNC), and Michigan State University (MSU).

This paper includes data collected by the TESS mission. Funding for the TESS mission is provided by the NASA Explorer Program. This work has made use of data from the European Space Agency (ESA) mission {\it Gaia} (\url{https://www.cosmos.esa.int/gaia}), processed by the {\it Gaia} Data Processing and Analysis Consortium (DPAC, \url{https://www.cosmos.esa.int/web/gaia/dpac/consortium}). Funding for the DPAC has been provided by national institutions, in particular the institutions participating in the {\it Gaia} Multilateral Agreement. This research has made use of the Exoplanet Follow-up Observation Program website, which is operated by the California Institute of Technology, under contract with the National Aeronautics and Space Administration under the Exoplanet Exploration Program. This work made use of the Washington Double Star Catalog maintained at USNO.

\vspace{1mm}

\facilities{SOAR (HRCam), TESS, Gaia}
\software{astropy \citep{astropy:2013, astropy:2018}, emcee \citep{emcee}, corner \citep{corner}}

\bibliography{refs.bib}

\appendix

\textbf{Notes for Table \ref{tab:ticbinaries}. -- }Column (1) is the TOI number. Columns (2) and (3) is the TIC number for the primary and secondary stars. Columns (4) to (6) gives the measured separation, position angle, and $I$-band contrast from the SOAR observations. Columns (7) to (9) give the separation and position angle for the system derived from the TIC coordinates, and the TESS band contrast for each pair of stars.

\footnotesize
\tabcolsep=0.12cm
\begin{longtable*}{ccc|ccc|ccc}
\caption{TIC matches to resolved binaries detected by SOAR \label{tab:ticbinaries}}\\
\hline
\hline
\noalign{\vskip 3pt}  
\text{TOI} & \text{TIC} & \text{TIC} & \multicolumn{3}{c}{SOAR}  & \multicolumn{3}{c}{TIC} \\ [0.1ex]
 &  Primary & Secondary & Sep. & P.A. & Contrast & Sep. & P.A. & Contrast  \\ [0.1ex]
  &  &  & (\arcsec) & ($^{\circ}$) & (mag)  & (\arcsec) & ($^{\circ}$) & (mag) \\ [0.1ex]
\hline
\noalign{\vskip 3pt}  
\endfirsthead
\multicolumn{9}{c}
{\tablename\ \thetable\ -- \textit{Continued}} \\
\hline \hline
\noalign{\vskip 3pt} 
\text{TOI} & \text{TIC} & \text{TIC} & \multicolumn{3}{c}{SOAR}  & \multicolumn{3}{c}{TIC} \\ [0.1ex]
 &  Primary & Secondary & Sep. & P.A. & Contrast & Sep. & P.A. & Contrast  \\ [0.1ex]
  &  &  & (\arcsec) & ($^{\circ}$) & (mag)  & (\arcsec) & ($^{\circ}$) & (mag) \\ [0.1ex]
\hline
\noalign{\vskip 3pt}  
\endhead
\endfoot
\hline
\endlastfoot
123 & 290131778 & 1992266045 & 1.2894 & 294.6 & 1.8 & 1.37 & 292.2 & 1.66 \\ 
128 & 391949880 & 675054894 & 2.2195 & 153.8 & 2.4 & 2.17 & 153.2 & 2.51 \\ 
147 & 220435095 & 685140266 & 2.6583 & 217.6 & 4.6 & 2.65 & 217.8 & 4.81 \\ 
149 & 260985861 & 675057530 & 1.117 & 132.6 & 0.1 & 0.73 & 132.0 & 0.82 \\ 
167 & 149990841 & 737110430 & 0.1593 & 306.6 & 0.8 & 0.39 & 126.0 & 0.06 \\ 
180 & 51912829 & 615712419 & 1.2682 & 63.7 & 4.3 & 1.69 & 60.8 & 4.15 \\ 
240 & 101948569 & 616347276 & 2.3656 & 197.6 & 3.1 & 2.3 & 197.7 & 3.19 \\ 
253 & 322063810 & 616169972 & 2.6623 & 351.2 & 4.0 & 2.58 & 353.9 & 3.82 \\ 
258 & 350445771 & 734530071 & 2.0749 & 343.2 & 2.9 & 2.07 & 343.5 & 2.93 \\ 
343 & 66497310 & 2052060639 & 2.8471 & 321.0 & 4.5 & 2.68 & 321.0 & 4.93 \\ 
386 & 238059180 & 767048826 & 1.1739 & 274.4 & 3.5 & 1.21 & 276.1 & 3.21 \\
387 & 92359850 & 651667037 & 2.2878 & 342.9 & 3.2 & 2.28 & 341.9 & 3.45 \\ 
427 & 70914192 & 686486697 & 2.6122 & 152.9 & 4.8 & 2.62 & 152.6 & 4.95 \\ 
433 & 188989177 & 188989178 & 4.0098 & 324.5 & 2.9 & 3.98 & 324.0 & 2.85 \\
498 & 121338379 & 803532221 & 2.1845 & 251.5 & 3.5 & 2.62 & 259.0 & 3.42 \\ 
594 & 146406806 & 824851862 & 2.8688 & 45.3 & 0.4 & 2.47 & 46.0 & 0.75 \\ 
611 & 154459165 & 831946954 & 2.3787 & 93.3 & 5.5 & 1.93 & 105.0 & 5.83 \\ 
637 & 133334108 & 821927265 & 2.3926 & 332.9 & 5.1 & 2.41 & 330.4 & 5.02 \\ 
645 & 157568289 & 157568287 & 4.0745 & 109.0 & 1.9 & 4.07 & 108.9 & 1.88 \\ 
651 & 293689267 & 293689266 & 3.2615 & 219.3 & 1.5 & 3.68 & 39.3 & -0.47 \\ 
658 & 48476907 & 48476908 & 3.88 & 67.2 & 0.7 & 3.87 & 68.4 & 0.71 \\ 
659 & 48476908 & 48476907 & 3.8774 & 67.2 & 1.0 & 3.87 & 248.4 & -0.71 \\ 
676 & 219187649 & 901927162 & 1.5023 & 260.6 & 1.5 & 1.5 & 261.2 & 1.9 \\ 
759 & 152147232 & 942050885 & 2.694 & 234.6 & 3.6 & 2.69 & 236.0 & 3.56 \\ 
772 & 286864983 & 951863424 & 2.3419 & 165.3 & 5.8 & 2.45 & 176.9 & 5.84 \\ 
837 & 460205581 & 847769574 & 2.3128 & 281.7 & 4.6 & 2.31 & 280.9 & 4.7 \\ 
851 & 40083958 & 610472559 & 1.8583 & 253.8 & 5.2 & 1.92 & 253.1 & 5.59 \\ 
847 & 231289421 & 650780404 & 1.2446 & 317.1 & 0.7 & 0.85 & 320.9 & 1.09 \\ 
905 & 261867566 & 1105898342 & 2.2757 & 100.8 & 5.9 & 2.14 & 101.5 & 6.21 \\ 
907 & 305424003 & 1510534043 & 3.6665 & 52.0 & 1.1 & 3.64 & 52.1 & 1.0

\end{longtable*}

\begin{figure*}
    \centering
    \includegraphics[scale=.65]{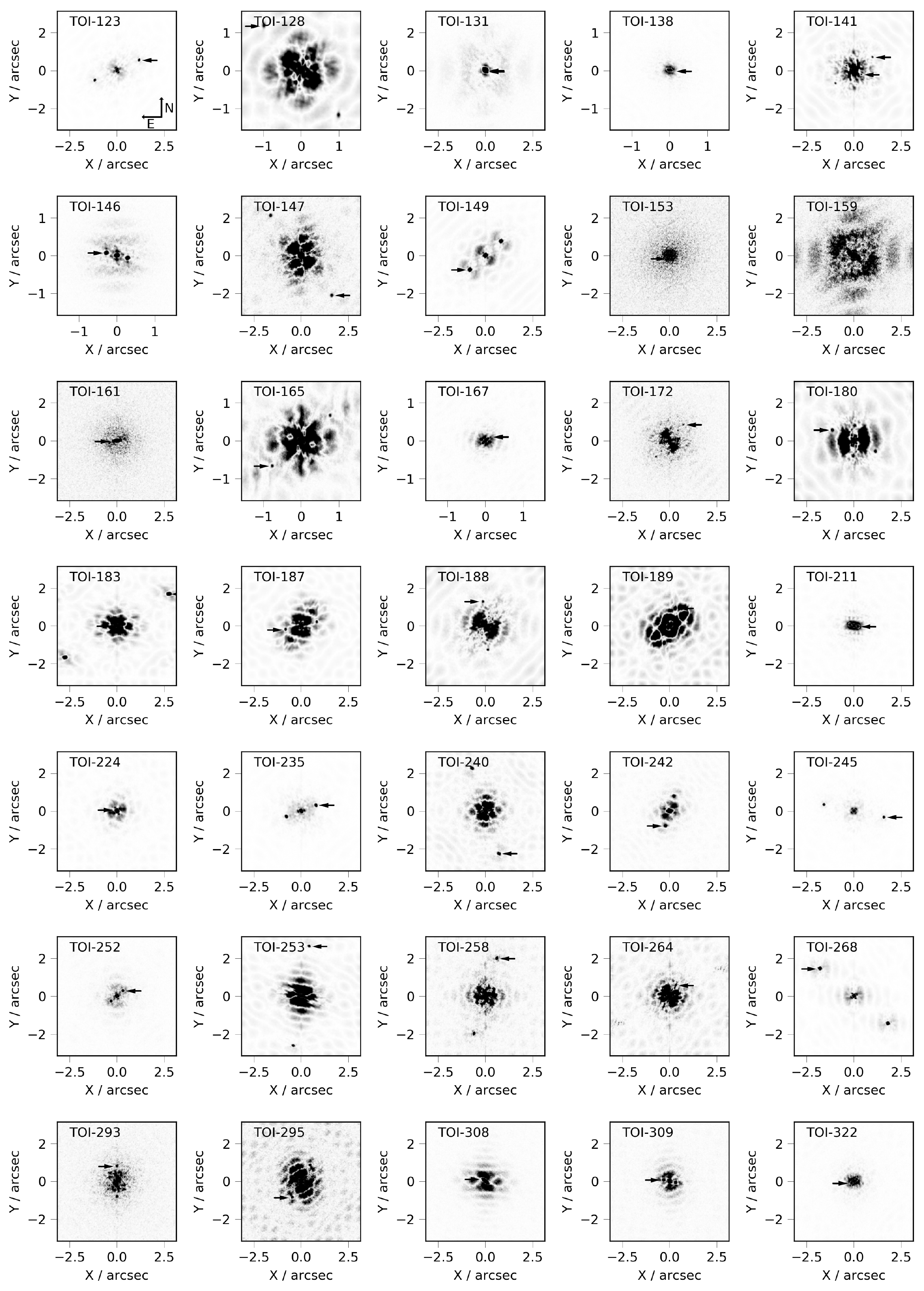}
    \caption{Speckle auto-correlation functions from SOAR speckle observing of TESS planet candidate hosts stars with resolved nearby stars. Each nearby star is mirrored in the images, with the true location marked by an arrow. Images are presented with an inverse linear scale for clarity. The orientation is similar in all images, with North pointed up and East to the left. A compass is shown in the top left image for reference.}
    \label{fig:grid1}
\end{figure*}

\begin{figure*}
    \centering
    \includegraphics[scale=.65]{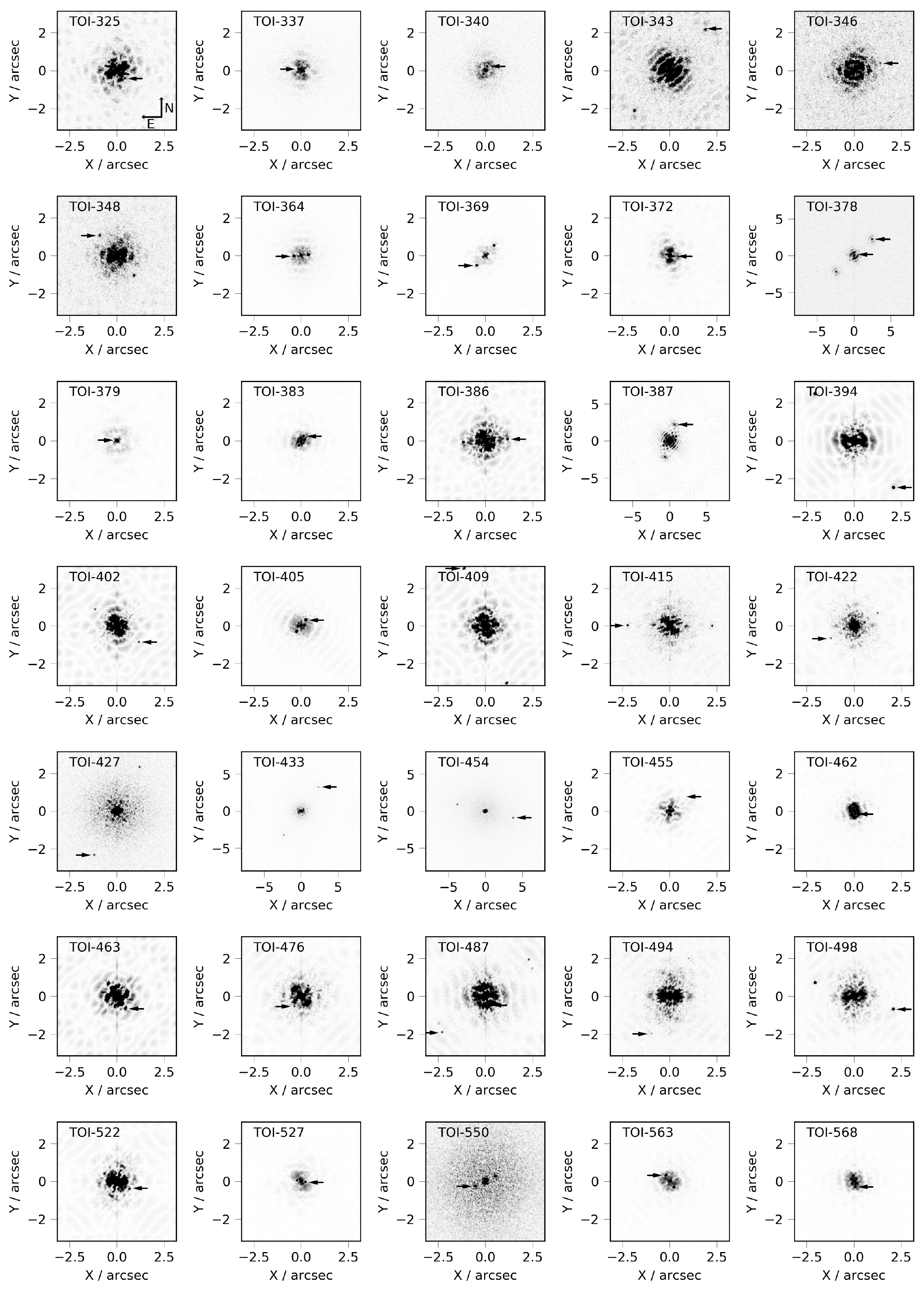}
    \caption{Similar to Figure \ref{fig:grid1}.}
    \label{fig:grid2}
\end{figure*}

\begin{figure*}
    \centering
    \includegraphics[scale=.65]{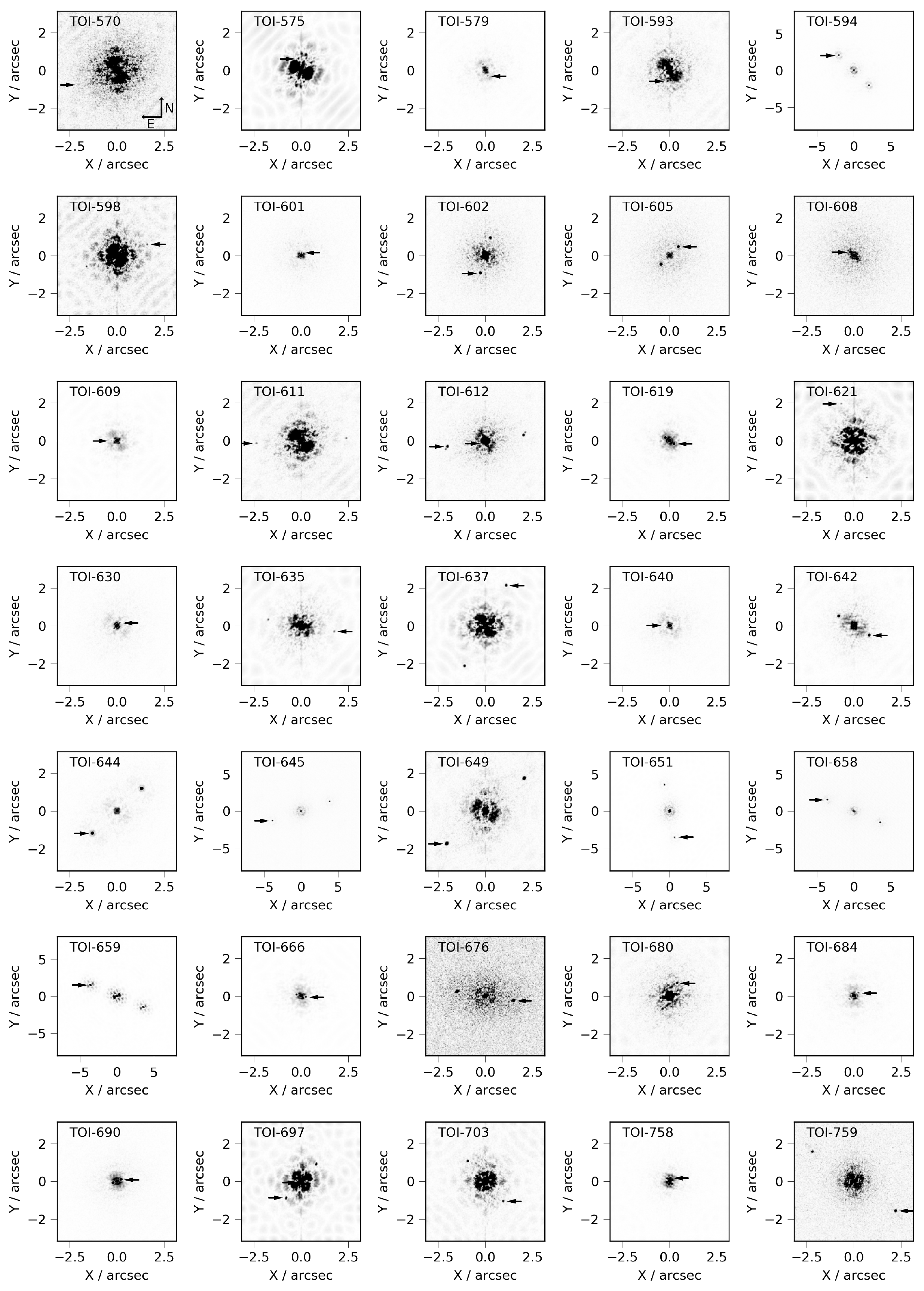}
    \caption{Similar to Figure \ref{fig:grid1}.}
    \label{fig:grid3}
\end{figure*}

\begin{figure*}
    \centering
    \includegraphics[scale=.65]{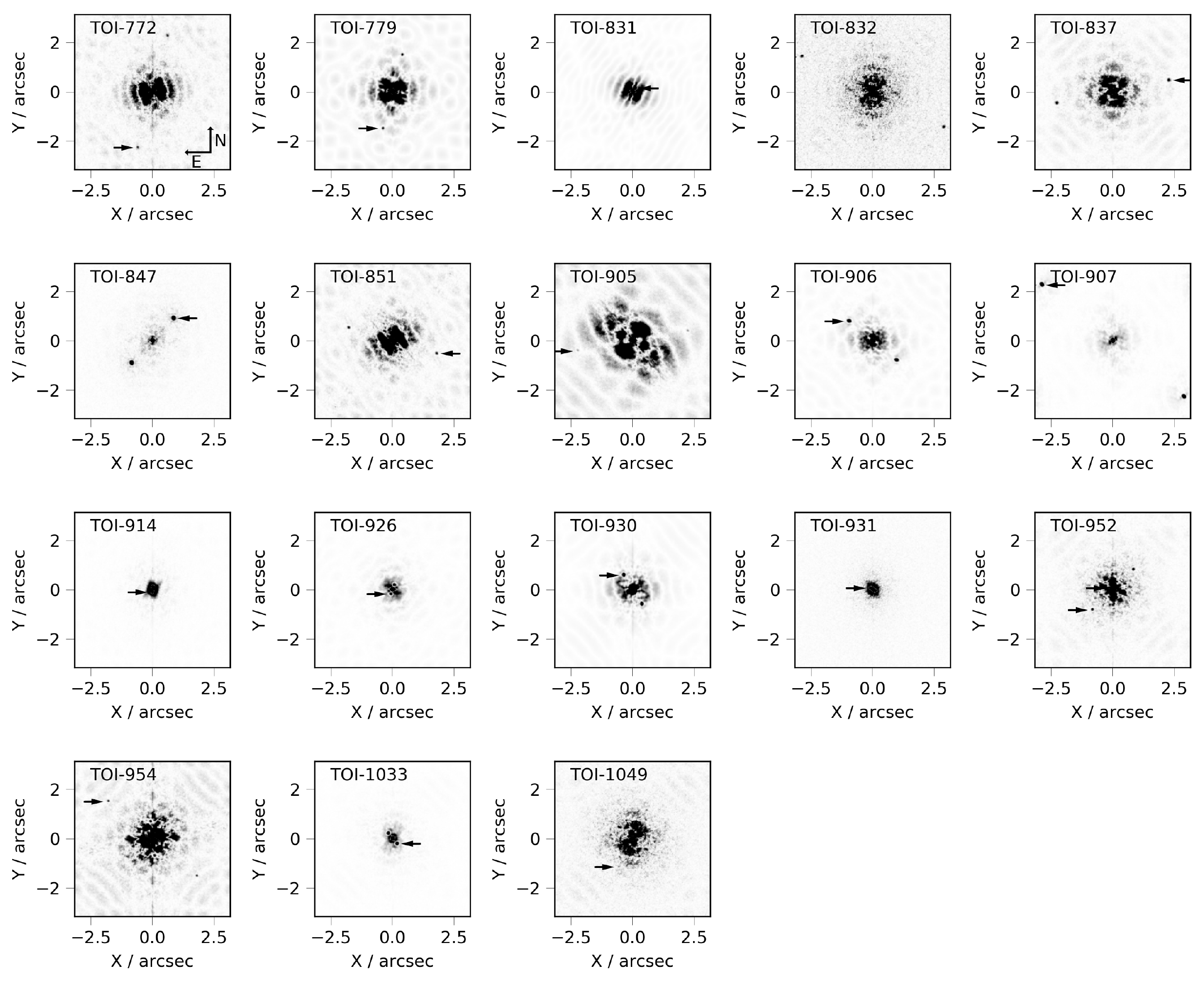}
    \caption{Similar to Figure \ref{fig:grid1}.}
    \label{fig:grid4}
\end{figure*}

\clearpage
\textbf{Notes for Table \ref{tab:binarybias}. -- }Column (1) is the TOI number. Columns (2) and (3) give the period and radius of the planet with the highest signal-to-noise ratio. Column (4) gives the stellar radius from the TIC of the primary star. Columns (5) and (6) give the transit duration and depth from the TESS light curves. Column (7) gives effective combined differential photometric precision, estimated from the the photometric precision model of \citet{tic} and interpolated to the duration of the transit. Column (8) is the number of TESS sectors the target was observed. Column (9) is the correction factor for the bias against planet detection in binary stars. The factor ranges from zero to one, with one indicating 100\% detection rate even with the flux contamination from a companion star.

\begin{longtable}{ccccccccc}
\caption{Detection Bias of TESS Planets in Multiple Stars\label{tab:binarybias}}\\
\hline
\hline
\noalign{\vskip 3pt}  
\text{TOI} & Period & R$_{P}$ & R$_{*}$ & Duration & Depth & CDPP$_{eff}$ & Sectors & $\alpha$ \\ [0.1ex]
 & (d) & (R$_{\oplus}$) & (R$_{\sun}$) & (hr) & (ppm) & (ppm) & & \\ [0.1ex]
\hline
\noalign{\vskip 3pt}  
\endfirsthead
\multicolumn{9}{c}
{\tablename\ \thetable\ -- \textit{Continued}} \\
\hline \hline
\noalign{\vskip 3pt} 
\text{TOI} & Period & R$_{P}$ & R$_{*}$ & Duration & Depth & CDPP$_{eff}$ & Sectors & $\alpha$ \\ [0.1ex]
 & (d) & (R$_{\oplus}$) & (R$_{\sun}$) & (hr) & (ppm) & (ppm) & & \\ [0.1ex]
\hline
\hline
\noalign{\vskip 3pt}  
\endhead
\endfoot
\hline
\endlastfoot
101 & 1.43 & 13.15 & 0.89 & 1.64 & 20452 & 597.0 & 1 & 1.0 \\ 
109 & 2.77 & 11.66 & 0.98 & 2.57 & 15691 & 822.2 & 1 & 1.0 \\ 
112 & 2.5 & 14.84 & 1.14 & 2.88 & 15044 & 277.6 & 4 & 1.0 \\ 
114 & 3.29 & 10.84 & 1.24 & 3.41 & 7125 & 148.2 & 6 & 1.0 \\ 
118 & 6.04 & 4.93 & 1.05 & 2.12 & 1704 & 97.0 & 1 & 1.0 \\ 
119 & 5.54 & 2.47 & 0.82 & 3.03 & 625 & 84.5 & 6 & 1.0 \\ 
120 & 11.54 & 12.22 & 2.34 & 8.56 & 2389 & 25.7 & 1 & 1.0 \\ 
121 & 14.78 & 14.78 & 1.23 & 5.45 & 14107 & 84.2 & 1 & 1.0 \\
\end{longtable}

\textbf{Notes for Table \ref{tab:gaiamatches}. -- }Columns (1) is the TOI number and Column (2) is the TIC number. Column (3) and (4) is the Gaia DR2 source ID for the primary and secondary stars. Columns (5) to (7) gives the measured separation, position angle, and $I$-band contrast from the SOAR observations. Columns (8) to (10) give the separation and position angle for the system derived from the Gaia DR2 coordinates, and the Gaia $G-$band contrast for each pair of stars.

\footnotesize
\tabcolsep=0.12cm
\begin{longtable*}{cccc|ccc|ccc}
\caption{Gaia DR2 matches to resolved binaries detected by SOAR \label{tab:gaiamatches}}\\
\hline
\hline
\noalign{\vskip 3pt}  
\text{TOI} & \text{TIC} & \multicolumn{2}{c}{Gaia DR2 IDs}& \multicolumn{3}{c}{SOAR}  & \multicolumn{3}{c}{Gaia DR2} \\ [0.1ex]
 & &  Primary & Secondary & Sep. & P.A. & Contrast & Sep. & P.A. & Contrast  \\ [0.1ex]
  &  & &  & (\arcsec) & ($^{\circ}$) & (mag)  & (\arcsec) & ($^{\circ}$) & (mag) \\ [0.1ex]
\hline
\noalign{\vskip 3pt}  
\endfirsthead
\multicolumn{10}{c}
{\tablename\ \thetable\ -- \textit{Continued}} \\
\hline \hline
\noalign{\vskip 3pt} 
\text{TOI} & \text{TIC} & \multicolumn{2}{c}{Gaia DR2 IDs}& \multicolumn{3}{c}{SOAR}  & \multicolumn{3}{c}{Gaia DR2} \\ [0.1ex]
 & &  Primary & Secondary & Sep. & P.A. & Contrast & Sep. & P.A. & Contrast  \\ [0.1ex]
  &  & &  & (\arcsec) & ($^{\circ}$) & (mag)  & (\arcsec) & ($^{\circ}$) & (mag) \\ [0.1ex]
\hline
\noalign{\vskip 3pt}  
\endhead
\endfoot
\hline
\endlastfoot
123 & 290131778 & 6790773586275235200 & 6790773586276283648 & 1.2894 & 294.6 & 1.8 & 1.299 & 294.18 & 1.74 \\ 
128 & 391949880 & 4621607053580909568 & 4621606297664620544 & 2.2195 & 153.8 & 2.4 & 2.2 & 153.78 & 2.55 \\ 
147 & 220435095 & 4762555785607165824 & 4762555785606163712 & 2.6583 & 217.6 & 4.6 & 2.642 & 217.78 & 4.73 \\ 
149 & 260985861 & 4621665155898448768 & 4621665155896340480 & 1.117 & 132.6 & 0.1 & 1.112 & 132.92 & 0.1 \\ 
165 & 350743714 & 4765618612685018368 & 4765618681404494976 & 2.4855 & 254.5 & 0.8 & 2.476 & 254.49 & 0.81 \\ 
180 & 51912829 & 4704412128965720448 & 4704412133261234176 & 1.2682 & 63.7 & 4.3 & 1.259 & 63.1 & 4.21 \\ 
240 & 101948569 & 4982951791883929472 & 4982951791883929600 & 2.3656 & 197.6 & 3.1 & 2.347 & 197.72 & 3.44 \\ 
245 & 154618248 & 6554931361478622976 & 6554931365773693952 & 1.6268 & 258.2 & 1.7 & 1.623 & 258.11 & 1.93 \\
\end{longtable*}

\begin{longrotatetable}
\textbf{Notes for Table \ref{tab:gaiabinaries}. -- }Column (1) is the TOI number and Column (2) is the TIC number. Columns (3) and (4) is the TIC number for the primary and secondary stars. Columns (5) and (6) gives the distance from the star from \citet{bailerjones18}. Columns (7) to (8) give the proper motion in Gaia DR2 for the primary and secondary stars. Column (9) gives the on-sky separation based on the Gaia DR2 coordinates, and Column (10) gives the projected physical separation using the average of the distances to the two stars. Column (11) gives the Gaia $G$-band contrast of the stellar pair.

\footnotesize
\tabcolsep=0.12cm

\begin{longtable*}{ccccccccccc}
\caption{Gaia DR2 binaries to TESS targets not detected by SOAR \label{tab:gaiabinaries}}\\
\hline
\hline
\noalign{\vskip 3pt}  
\text{TOI} & \text{TIC} & \multicolumn{2}{c}{Gaia DR2 ID}  & \multicolumn{2}{c}{Distance} & \multicolumn{2}{c}{Proper motion} & \text{Sep.} & \text{Projected sep.} & \text{Gaia contrast} \\ [0.1ex]
 &  & Primary & Secondary & Primary & Secondary & Primary & Secondary &  &  &  \\ [0.1ex]
 \hline
  &  &  &  & (pc) & (pc)  & (mas/yr)  & (mas/yr)  & (\arcsec) & (AU) & (mag) \\ [0.1ex]
\hline
\noalign{\vskip 3pt}  
\endfirsthead
\multicolumn{11}{c}
{\tablename\ \thetable\ -- \textit{Continued}} \\
\hline \hline
\noalign{\vskip 3pt} 
\text{TOI} & \text{TIC} & \multicolumn{2}{c}{Gaia DR2 ID}  & \multicolumn{2}{c}{Distance} & \multicolumn{2}{c}{Proper motion} & \text{Sep.} & \text{Projected sep.} & \text{Gaia contrast} \\ [0.1ex]
 &  & Primary & Secondary & Primary & Secondary & Primary & Secondary & (\arcsec) & (AU) & (mag) \\ [0.1ex]
\hline
\noalign{\vskip 3pt}  
\endhead
\endfoot
\hline
\endlastfoot
129 & 201248411 & 4923860051276772608 & 4923860051276772480 & 61.8 & 61.9 & 215.1 & 219.7 & 3.81 & 235.5 & 4.8 \\ 
130 & 263003176 & 4617759514501503616 & 4617759518796452608 & 57.6 & 56.1 & 142.7 & 151.4 & 2.25 & 129.6 & 5.42 \\ 
143 & 25375553 & 6813902839862151936 & 6813902839862983296 & 298.1 & 293.1 & 13.6 & 14.3 & 5.03 & 1499.4 & 5.02 \\ 
174 & 425997655 & 4674216245427964416 & 4674216245427964672 & 39.0 & 39.2 & 111.6 & 111.5 & 9.57 & 373.2 & 6.45 \\ 
183 & 183979262 & 4686544382117423488 & 4686544382117422720 & 213.0 & 215.4 & 79.4 & 80.4 & 3.21 & 683.7 & 2.63 \\ 
199 & 309792357 & 4762582895440787712 & 4762582861080613248 & 102.3 & 101.8 & 74.2 & 73.4 & 9.99 & 1022.0 & 7.84 \\ 
200 & 410214986 & 6387058411482257536 & 6387058411482257280 & 44.1 & 44.1 & 104.2 & 102.0 & 5.36 & 236.4 & 1.08 \\ 
204 & 281781375 & 4903786336207800576 & 4903786336207800704 & 95.3 & 95.1 & 173.5 & 174.0 & 10.25 & 976.8 & 1.16 \\ 
222 & 144440290 & 6531037981670835584 & 6531037981670835456 & 84.0 & 81.1 & 178.5 & 177.4 & 3.46 & 290.6 & 8.02 \\ 
229 & 120610833 & 2308834780352875904 & 2308834784647991168 & 211.2 & 233.7 & 70.7 & 70.5 & 4.37 & 922.9 & 6.96 \\ 
248 & 201793781 & 4743138925656526976 & 4743138925656526848 & 76.0 & 75.7 & 94.4 & 95.6 & 5.02 & 381.5 & 5.45 \\ 
268 & 219253008 & 4779340346001388160 & 4779341750455118720 & 320.5 & 325.9 & 21.5 & 21.1 & 9.99 & 3201.8 & 8.13 \\ 
277 & 439456714 & 2353440974955205504 & 2353440974955205632 & 64.7 & 64.7 & 273.5 & 274.6 & 17.25 & 1116.1 & 0.3 \\ 
300 & 166697854 & 4743406244421179264 & 4743406244421179136 & 397.4 & 377.7 & 15.2 & 16.0 & 4.1 & 1629.3 & 4.66 \\ 
354 & 100097716 & 4959658534969336320 & 4959658534969336576 & 366.8 & 404.5 & 28.6 & 28.5 & 4.84 & 1775.3 & 0.65 \\ 
354 & 299780329 & 4632142191744789888 & 4632142196040368896 & 305.6 & 306.9 & 17.2 & 17.2 & 7.37 & 2252.3 & 3.25 \\ 
368 & 77031413 & 2323985539482908416 & 2323985535188372480 & 233.0 & 233.6 & 88.4 & 87.9 & 6.12 & 1426.0 & 0.59 \\ 
377 & 139285736 & 6529468772420035712 & 6529468772420035584 & 244.3 & 259.0 & 12.0 & 10.4 & 9.29 & 2269.5 & 3.14 \\ 
381 & 207084429 & 4917040330405933824 & 4917040330405933952 & 75.5 & 82.3 & 152.9 & 156.8 & 4.94 & 373.0 & 7.62 \\ 
390 & 250386181 & 2491782421314919040 & 2491782417019596160 & 167.3 & 171.3 & 38.4 & 38.2 & 6.28 & 1050.6 & 6.32 \\ 
393 & 29960109 & 5157183324996790272 & 5157183324996789760 & 37.9 & 37.9 & 170.6 & 180.2 & 8.35 & 316.5 & 0.29 \\ 
396 & 178155732 & 5064574720469473792 & 5064574724768583168 & 32.0 & 30.9 & 162.4 & 173.2 & 8.36 & 267.5 & 9.8 \\ 
397 & 219379012 & 4785828357959075840 & 4785828357959076480 & 167.0 & 168.4 & 31.3 & 29.8 & 4.32 & 721.4 & 4.84 \\ 
412 & 7624182 & 4844367147295252864 & 4844365669826503808 & 460.7 & 446.6 & 11.5 & 11.3 & 7.88 & 3630.3 & 6.0 \\ 
414 & 325680697 & 5171630735987857152 & 5171630735987104512 & 115.1 & 114.8 & 48.3 & 48.6 & 4.5 & 517.9 & 5.25 \\ 
426 & 189013224 & 2983316311375470976 & 2983316311375257472 & 113.6 & 112.6 & 26.2 & 27.8 & 8.88 & 1008.8 & 0.77 \\ 
440 & 143350972 & 2971536418673198976 & 2971536418671276544 & 49.3 & 49.1 & 113.0 & 112.1 & 6.45 & 318.0 & 4.87 \\ 
441 & 316916655 & 2987545475475708416 & 2987545548491503360 & 184.0 & 177.7 & 10.9 & 12.3 & 15.64 & 2877.8 & 6.0 \\ 
470 & 37770169 & 2912264564319611136 & 2912264564316598272 & 130.5 & 130.1 & 85.1 & 84.4 & 13.6 & 1774.8 & 6.25 \\ 
470 & 37770169 & 2912264564319611136 & 2912264564316598144 & 130.5 & 130.5 & 85.1 & 86.4 & 13.74 & 1793.1 & 5.89 \\ 
498 & 121338379 & 3071584413361857280 & 3071584417656748928 & 188.4 & 176.6 & 23.4 & 21.7 & 5.2 & 979.7 & 7.25 \\ 
505 & 268644785 & 5489780919480009472 & 5489780919477609600 & 324.7 & 315.8 & 5.1 & 5.4 & 6.09 & 1977.4 & 6.65 \\ 
510 & 238086647 & 5508330367833229824 & 5508330367831469952 & 93.1 & 94.0 & 124.6 & 122.8 & 5.48 & 510.2 & 3.93 \\ 
575 & 386435344 & 5709861464697757824 & 5709861468994311424 & 182.0 & 156.4 & 48.6 & 48.5 & 5.36 & 975.5 & 8.28 \\ 
580 & 81419525 & 5519619186857962112 & 5519619191155977344 & 346.5 & 475.4 & 11.6 & 9.1 & 14.09 & 4882.2 & 10.99 \\ 
581 & 180987952 & 5525188767305211904 & 5525188728649771392 & 434.2 & 428.8 & 13.6 & 20.8 & 7.8 & 3386.8 & 8.56 \\ 
592 & 196286587 & 5695996352497664512 & 5695996348197148032 & 358.4 & 385.6 & 8.6 & 33.0 & 11.28 & 4042.8 & 1.49 \\ 
638 & 78154865 & 3822912388299321728 & 3822912392594624128 & 96.6 & 97.6 & 14.3 & 16.0 & 10.53 & 1017.2 & 4.32 \\ 
648 & 78672342 & 2933433564771909888 & 2933433564771909760 & 637.3 & 814.6 & 11.0 & 11.6 & 2.95 & 1880.0 & 5.9 \\ 
650 & 349373192 & 5293295919554813952 & 5293296130007438848 & 402.8 & 348.7 & 12.6 & 12.9 & 8.97 & 3613.1 & 8.93 \\ 
670 & 147660201 & 5388433503908280320 & 5388433503908279168 & 191.5 & 183.6 & 57.9 & 56.4 & 13.79 & 2640.8 & 1.91 \\ 
708 & 391821647 & 4657888149862830720 & 4657888150116259584 & 214.8 & 347.2 & 24.0 & 21.4 & 13.74 & 2951.4 & 9.36 \\ 
756 & 73649615 & 6129327525817451648 & 6129327319659021056 & 86.2 & 86.2 & 218.3 & 217.6 & 11.09 & 956.0 & 1.35 \\ 
764 & 181159386 & 5385762893242980992 & 5385762824520652160 & 789.3 & 754.5 & 6.5 & 6.4 & 5.96 & 4704.2 & 6.21 \\ 
811 & 100757807 & 2890519660294945920 & 2890519660294833792 & 284.0 & 283.8 & 29.7 & 29.1 & 4.33 & 1229.7 & 1.83 \\  
815 & 102840239 & 5415648821879172096 & 5415648821874435584 & 59.7 & 59.3 & 11.4 & 12.3 & 6.04 & 360.6 & 2.48 \\ 
824 & 193641523 & 5880886001621564928 & 5880886001577333888 & 63.9 & 69.2 & 160.9 & 3.3 & 6.7 & 428.1 & 9.16 \\ 
829 & 276128561 & 6199033466340798464 & 6199033466344807040 & 142.5 & 138.8 & 71.0 & 71.4 & 5.46 & 778.0 & 5.28 \\ 
832 & 350332997 & 4768613025228556416 & 4768613029524059648 & 586.7 & 617.2 & 13.7 & 12.9 & 3.7 & 2170.8 & 4.76 \\ 
858 & 198008005 & 4683737294568479104 & 4683737294569921664 & 255.8 & 250.5 & 15.1 & 15.5 & 10.95 & 2801.0 & 0.28 \\ 
878 & 219380235 & 4771537902252716416 & 4771537897957355392 & 334.5 & 324.6 & 35.5 & 34.9 & 5.64 & 1886.6 & 6.49 \\ 
915 & 259389219 & 4781316164099781760 & 4781316168396825728 & 506.4 & 443.0 & 10.1 & 29.7 & 6.05 & 3063.7 & 4.18 \\ 
938 & 332660150 & 3189432444744896512 & 3189432444743495936 & 215.0 & 195.0 & 12.6 & 12.9 & 7.95 & 1709.2 & 7.6 \\ 
\end{longtable*}
\end{longrotatetable}
\clearpage

\begin{longrotatetable}
\textbf{Notes for Table \ref{tab:whitelist}. -- }Columns (1) and (2) give the TOI and TIC numbers, respectively. Column (3) designates the components of the resolved binaries according to the WDS style (mostly 'AB'). This matters for resolved triple systems, indicating their hierarchy. The equatorial coordinates for J2000, in degrees, are given in Columns (4) and (5). Column (6) give the filter (mostly $I$, with a few targets observed also in $V$), Column (7) the date of the observation (in Julian years). For resolved binaries, Columns (8) and (10) give the position angle $\theta$ and the separation $\rho$, while Columns (9) and (11) contain estimates of the measurement errors in tangential ($\rho \sigma_\theta$) and radial ($\sigma_\rho$) directions, in mas. The measured magnitude difference $\Delta m$ is given on Column (12). Some targets have multiple measurements. For unresolved sources (single stars), the Columns (8) to (12) are empty. Flags for the photometry are provided in Column (13). These flags are ':' for a companion with a low signal-to-noise ratio, 'q' for an identified quadrant from the shift-and-add images, '*' if the photometry is corrected for anisoplanatism using the average image. The estimated resolution limit is listed in Column (14) for all stars; Columns (15) and (16) give the estimated maximum detectable $\Delta m$ at separations of 0\farcs15 and 1\arcsec.

\tiny
\tabcolsep=0.2cm
\begin{longtable}{cccccccccccccccc}
\caption{Full SOAR Speckle Observation List and Binary Properties \label{tab:whitelist}}\\
\hline
\hline
\noalign{\vskip 3pt}  
\text{TOI} & TIC & Comp. & \multicolumn{2}{c}{R.A., Dec. (J2000)} & Filter & \text{Obs.} & $\theta$ & $\rho*$$\sigma_{\theta}$ & $\rho$ & $\sigma_{\rho}$ & $\Delta m$ & Flag & Min. $\rho$ & \multicolumn{2}{c}{Limiting $\Delta m$} \\ [0.1ex]
 &  & & (deg) & (deg) &  & Year & ($^{\circ}$) & (mas) & (\arcsec) & (mas) & (mag) & & (\arcsec) & 0\farcs15 & 1\arcsec \\ [0.1ex]
\hline
\noalign{\vskip 3pt}  
\endfirsthead
\multicolumn{16}{c}
{\tablename\ \thetable\ -- \textit{Continued}} \\
\hline \hline
\noalign{\vskip 3pt} 
\text{TOI} & TIC & Comp. & \multicolumn{2}{c}{R.A., Dec. (J2000)} & Filter & \text{Obs.} & $\theta$ & $\rho*$$\sigma_{\theta}$ & $\rho$ & $\sigma_{\rho}$ & $\Delta m$ & Flag & Min. $\rho$ & \multicolumn{2}{c}{Limiting $\Delta m$} \\ [0.1ex]
 &  & & (deg) & (deg) &  & Year & ($^{\circ}$) & (mas) & (\arcsec) & (mas) & (mag) & & (\arcsec) & 0\farcs15 & 1\arcsec \\ [0.1ex]
\hline
\hline
\noalign{\vskip 3pt}  
\endhead
\endfoot
\hline
\endlastfoot

101 & 231663901 & --- & 318.73713 & -55.87186 & I & 2018.731 &  &  &  &  &  &  & 0.067 & 2.20 & 2.97 \\ 
109 & 29344935 & --- & 313.21605 & -25.68764 & I & 2019.375 &  &  &  &  &  & : & 0.098 & 0.00 & 0.00 \\ 
112 & 388104525 & --- & 55.93345 & -65.19383 & I & 2019.534 &  &  &  &  &  &  & 0.064 & 2.33 & 3.88 \\ 
114 & 25155310 & --- & 63.37440 & -69.22675 & I & 2019.534 &  &  &  &  &  &  & 0.064 & 2.51 & 4.37 \\ 
118 & 266980320 & --- & 349.56216 & -56.90408 & I & 2018.731 &  &  &  &  &  &  & 0.048 & 3.00 & 4.05 \\ 
119 & 278683844 & --- & 99.23777 & -58.01477 & I & 2018.732 &  &  &  &  &  &  & 0.048 & 2.36 & 3.72 \\ 
120 & 394137592 & --- & 4.44608 & -66.35889 & I & 2018.731 &  &  &  &  &  &  & 0.043 & 2.57 & 3.82 \\ 
121 & 207081058 & --- & 331.86741 & -41.81472 & I & 2018.731 &  &  &  &  &  &  & 0.052 & 2.33 & 3.67 \\
\end{longtable}

\end{longrotatetable}

\end{document}